\documentclass[twocolumn,eqsecnum,showpacs,amsmath,amssymb,superscriptaddress,nofootinbib,english,datedpages]{revtex4}

\usepackage{babel}

\begin{document}

\title{Higher-dimensional Kundt waves and gyratons}

\author{Pavel Krtou\v{s}}%
\email[e-mail: ]{Pavel.Krtous@utf.mff.cuni.cz}
 \affiliation{%
 Institute of Theoretical Physics, Faculty of Mathematics and Physics, Charles University,\\
  V Hole\v{s}ovi\v{c}k\'{a}ch 2, 180 00 Prague 8, Czech Republic
}%
\author{Ji\v{r}\'{\i} Podolsk\'{y}}%
\email[e-mail: ]{podolsky@mbox.troja.mff.cuni.cz}
 \affiliation{%
 Institute of Theoretical Physics, Faculty of Mathematics and Physics, Charles University,\\
  V Hole\v{s}ovi\v{c}k\'{a}ch 2, 180 00 Prague 8, Czech Republic
  }%
\author{ Andrei Zelnikov}%
 \email[e-mail: ]{zelnikov@ualberta.ca}
\affiliation{%
Theoretical Physics Institute, Department of Physics, University of Alberta,\\
Edmonton, Alberta, Canada T6G 2E1
}%
\author{Hedvika Kadlecov\'{a}}%
\email[e-mail: ]{hedvika.kadlecova@centrum.cz}
 \affiliation{%
 Institute of Theoretical Physics, Faculty of Mathematics and Physics, Charles University,\\
  V Hole\v{s}ovi\v{c}k\'{a}ch 2, 180 00 Prague 8, Czech Republic
  }%

\date{January 13, 2012} 

\begin{abstract}

We present and analyze exact solutions of the Einstein--Maxwell equations in higher dimensions which form a large subclass of the Kundt family of spacetimes. We assume that the cosmological constant may be nonvanishing, and the matter consists of a background aligned electromagnetic field and an additional pure radiation (gyratonic) field with a spin. We show that the field equations reduce to a set of linear equations on the transverse space which can be solved exactly and expressed in terms of the Green functions. We thus find explicit exact gyratonic gravitational and electromagnetic fields created by a radiation beam of null
matter with arbitrary profiles of energy density and angular momenta. In the absence of the gyratonic matter we obtain pure nonexpanding higher-dimensional gravitational waves.

In particular, we investigate gyratons and waves propagating on backgrounds which are a direct-product of 2-spaces of constant curvature. Such type D or 0 background spacetimes generalize 4-dimensional Nariai, anti-Nariai and Pleba\'{n}ski--Hacyan universes, and conformally flat Bertotti--Robinson and Minkowski spaces. These spacetimes belong to a wider class of spaces which admit the K\"ahler structure related to the background magnetic field. The obtained wave and gyraton solutions are also members of the recently discussed class of spacetimes with constant scalar invariants (CSI) of the curvature tensor.

\end{abstract}

\maketitle


\section{Introduction}\label{sc:intro}

Due to the nonlinear nature of the Einstein equations,
finding their exact solutions has always been an important and challenging
problem. It is a pleasant surprise
if one can find an exact solution of such nonlinear equations, especially when
they describe the gravitational field of a physically meaningful gravitational
source.

In this paper, we study gyratons, a particularly interesting class of solutions of the
Einstein--Maxwell equations in four and higher dimensions which describe
the gravitational and electromagnetic fields of a pulse of a circularly polarized
radiation beam or spinning ultrarelativistic particles.
Generally speaking the gyraton solutions represent the
fields of a localized matter source with an intrinsic rotation which
is moving at the speed of light.

Historically, the gravitational fields generated by
light pulses without spin were originally studied in linear approximation by Tolman
\cite{Tolman:1934} in 1934. Later, exact solutions of the Einstein--Maxwell equations for such
`pencils of light' were found and analyzed by Peres \cite{Peres:1960}
and Bonnor \cite{Bonnor:1969a,Bonnor:1969b,Bonnor:1970a}.
These solutions belong to a general family of pp-waves
\cite{Brinkmann:1925,Stephani:2003,GriffithsPodolsky:2009}. The simplest of such solutions
describes the gravitational impulsive wave created by an infinitely thin beam
with the delta-like time distribution of the light-pulse.
This is the famous Aichelburg--Sexl metric \cite{AichelburgSexl:1971} which
represents the field of a point-like null particle without a spin.
It can be derived from the Schwarzschild black hole metric in the
ultrarelativistic limit,
by taking the distributional limit of exact sandwich waves, or by the Penrose
`cut and paste' approach. The review of more general solutions for such impulsive waves
\cite{FerrariPendenza:1990,LoustoSanchez:1992,HottaTanaka:1993,
BalasinNachbagauer:1995,BalasinNachbagauer:1996,PodolskyGriffiths:1997,
PodolskyGriffiths:1998} and their construction methods can be found in
\cite{Podolsky:2002,BarrabesHogan:2003}. Gravitational field of a spinning
null fluid---so called spinning nullicon---was first discussed in 1970 by
Bonnor \cite{Bonnor:1970b}. Possible interpretation of such matter source
as a massless neutrino fields was given by Griffiths \cite{Griffiths:1972}.
Recently, these solutions have been investigated in greater detail and generalized
by Frolov, Israel, Fursaev and Zelnikov, who have called them gyratons
\cite{FrolovFursaev:2005,FrolovIsraelZelnikov:2005}.

In four dimensions, the gyratons propagating on the
Minkowski background are, outside the source, locally isometric to standard pp-waves. Using a suitable
gauge transformation one can always set to zero the nondiagonal terms in the gyraton
metric, namely the $g_{ui}$ components in the Brinkmann form \cite{Brinkmann:1925},
which reflect the rotational part of the metric. However, they can not be
eliminated globally, because the gauge invariant contour integral $\oint
g_{ui}(u,x^i)\,dx^i$ around the position of the gyraton is proportional to the
angular momentum density of the gravitating source, which is generally nonvanishing.

In higher dimensions, the structure of the source and its gravitational field is
much richer. Thus, in a generic case, outside the source one can not make
the nondiagonal metric components $g_{ui}$ zero, even locally. Such higher-dimensional
gyratons propagating in an asymptotically flat ${D}$-dimensional spacetime were
studied in \cite{FrolovIsraelZelnikov:2005}. It was demonstrated that the Einstein
equations for gyratons reduce to a set of linear equations in the Euclidean
$(D{-}2)$-dimensional transverse space, and that the gyraton
metrics belong to a class of so called VSI spacetimes.\footnote{%
In VSI spacetimes all polynomial scalar invariants, constructed from
the curvature tensor and its covariant derivatives, vanish identically \cite{PravdaPravdovaColeyMilson:2002}.
For the discussion of spacetimes with nonvanishing but nonpolynomial
scalar curvature invariants, see \cite{Page:2009}.}
These spacetimes have unique mathematical features. For example, due to their remarkable
geometrical properties they describe not only the exact classical solution
for the gyraton but also the exact solution of a corresponding quantum
problem. This happens because quantum corrections to the
classical solution vanish identically in all loops
\cite{ColeyGibbonsHervikPope:2008}.

Subsequently, there were found various generalizations of the gyratons:
charged ultrarelativistic sources propagating on the background of $D$-dimensional
Minkowski space \cite{FrolovZelnikov:2006}, gyratons in the
asymptotically anti-de Sitter spacetime \cite{FrolovZelnikov:2005},
and gyratons in the Melvin spacetime \cite{KadlecovaKrtous:2010}.
In the case of anti-de~Sitter background the obtained gyratons generalize the Siklos
family of nonexpanding waves with a negative cosmological constant \cite{Siklos:1985}
(studied in detail in \cite{Podolsky:1998}) which belong to the class of CSI spacetimes
with constant scalar invariants
\cite{ColeyHervikPelavas:2006,ColeyGibbonsHervikPope:2008,
ColeyHervikPelavas:2008,ColeyHervikPelavas:2009}.
Supersymmetric gyratons, as solutions of  minimal gauged
supergravity in five dimensions, has been studied in
\cite{CaldarelliKlemmZorzan:2007}. The solutions of supergravity equations
generated by string-like sources moving with the
speed of light, the string gyratons, have been found in \cite{FrolovLin:2006}.

In our recent paper \cite{KadlecovaZelnikovKrtousPodolsky:2009} a large new
class of 4-dimensional gyratons on the direct-product spacetimes has been found.
We have shown that this class of gyratons has properties similar to
other gyratonic solutions: the Einstein equations reduce to a set of
linear differential equations in the transverse constant-curvature 2-space,
and these spacetimes also belong to the CSI class.
Our results are applicable to the description of
gyratons in the vicinity of the horizon of extremely
charged black holes. Indeed, the near-horizon geometry of these
background spacetimes is $AdS_2\times S_2$
and belongs exactly to the type of spacetimes we have considered.

In fact, all the spaces studied in \cite{KadlecovaZelnikovKrtousPodolsky:2009}
belong to the family of Kundt spacetimes defined as those admitting a null vector
field that is geodesic, without expansion, shear and twist
\cite{Stephani:2003,GriffithsPodolsky:2009}. The Kundt spacetimes are of
great importance in standard general relativity and recently they have found
a number of interesting applications in higher-dimensional theories,
namely because of their unique curvature and holonomy structure.
Moreover, the Kundt class involves various important special cases such as the pp-waves,
VSI and CSI spacetimes \cite{PodolskyZofka:2009,ColeyEtal:2009}.

In this paper, we are going to study higher-dimensional gyratons in the Kundt family in detail,
generalizing thus our results obtained in \cite{KadlecovaZelnikovKrtousPodolsky:2009}
to the case of direct-product background spacetimes of arbitrary dimensions.

In order to find sufficiently general direct-product
spacetimes one need to include an additional background electromagnetic field.
Therefore, we first solve the Einstein--Maxwell equations for such backgrounds.
When we subsequently add a null matter source with angular momenta to the system, both the
metric and the electromagnetic field is deformed so that the structure of
the spacetime will not be of a direct-product type anymore. It is thus a highly
nontrivial fact that the complete nonlinear Einstein--Maxwell equations can still be
reduced to a set of linear differential equations formulated on the background transverse space.
The nonlinearity boils down to the nonlinear dependence of the sources on the
right-hand sides of the nonhomogeneous differential equations for the rotational components
${F_{ui}}$ and ${g_{ui}}$ of the electromagnetic field and the metric, respectively.
As a result, we are able to express explicitly the corresponding exact solutions of the Einstein--Maxwell
equations coupled to the gyratonic matter in terms of the well-known Green functions.

The structure of our paper is as follows.
The following section briefly reviews the geometrical construction of a
general Kundt metric in any dimension and introduces its naturally adapted
coordinates and matter content. In Sec.~\ref{sc:trsp} an important method
of splitting geometrical quantities into the temporal and transverse directions
is introduced. The main goal there is to formulate all physical equations
on the ${(D{-}2)}$-dimensional transverse space, which is done in Sec.~\ref{sc:fieldeq}.
The strategy of solving the field equations, including the classification of their solutions,
is discussed in Sec.~\ref{sc:disc}. The most interesting solutions are then investigated in Sec.~\ref{sc:solvable}. We explicitly decouple the field equations for spacetimes with
a vanishing magnetic field, and also for gyratons on direct-product background
spacetimes with uniform electric and magnetic fields.
The character and significance of these backgrounds is discussed.
The paper concludes with a short summary and Appendix~\ref{apx:gauge} describing the gauge freedom of the
Kundt metric parametrization.


\section{Metric and matter}\label{sc:ansatz}

\subsection*{Spacetime geometry}

The Kundt class consists of spacetimes which admit a \emph{nontwisting, nonexpanding, and shear-free geodesic congruence} generated by a null vector field $\tens{k}$. This field~${\tens{k}}$ also generates a family of null hypersufaces which forms a foliation ${S}$ of the whole spacetime.
Each of these ${(D{-}1)}$-dimensional null hypersurfaces can further be foliated by ${(D{-}2)}$-dimensional spatial \emph{transverse spaces}.

In the present work we restrict our attention to a specific subclass of the general Kundt family. Namely, we assume that the 2-spaces of vectors orthogonal to the transverse spaces are \emph{integrable} (see below for more technical details).

Under these geometric assumptions the spacetime metric ${\tens{g}}$ can be written in the form\footnote{%
We omit tensor-product symbol~${\otimes}$, e.g., ${\stgrad u\, \stgrad u = \stgrad u \otimes \stgrad u}$. We use the standard convention for the wedge product ${\tens{\alpha}\wedge\tens{\beta} = \tens{\alpha}\,\tens{\beta}-\tens{\beta}\,\tens{\alpha}}$. Analogously, we denote by `${\vee}$' the symmetrical tensor product, for example, ${\stgrad u \vee \ta = \stgrad u\, \ta + \ta\, \stgrad u}$.
Equivalent expressions for nontrivial metric components ${g_{\mu\nu}}$ are
${g_{uu} = -2H}$,
${g_{ur} = -1}$,
${g_{ui} = a_i}$,
${g_{ij} = q_{ij}}$.
The components ${g^{\mu\nu}}$ of the inverse metric are thus
${g^{rr} = 2H + a^2}$,
${g^{ur} = -1}$,
${g^{ri} = a^i}$,
${g^{ij} = q^{ij}}$.
}
\cite{Stephani:2003,GriffithsPodolsky:2009,PodolskyZofka:2009,ColeyEtal:2009,KrtousPodolsky:inprep}
\begin{equation}\label{metric}
  \tens{g} = - 2 H\, \stgrad u\, \stgrad u - \stgrad u \vee \stgrad r + \stgrad u \vee \ta + \mtr\;,
\end{equation}
where ${H}$ is a scalar function, ${\ta}$ is a transverse 1-form, which we will call the \emph{metric 1-form}, and ${\mtr}$ is the \emph{transverse metric}, metric on the transverse space.

To distinguish the spacetime quantities from those in the Riemannian transverse space we use the superscript `${{}^\st}$' on the left of spacetime objects or operations. For example, `${\stgrad}$' is the spacetime gradient and external derivative, and `${\stcovd\,}$' denotes the spacetime covariant derivative associated with the metric ${\tens{g}}$.

Let us briefly comment on the geometrical meaning of the coordinates and other quantities introduced in the metric~\eqref{metric}.
The coordinate ${u}$ is adjusted to the null foliation ${S}$. We denote ${S_u}$ the specific hypersurface of this foliation corresponding to a given constant value of~${u}$. The null generator ${\tens{k}}$ is tangent to ${S}$, and it is normalized in accordance with the ${u}$-coordinate as\footnote{%
The central dot `${\cdot}$' denotes the contraction, i.e., $\tens{k}\cdot\stgrad u=k^\mu\,{}^\st\! d_\mu u=k^\mu u_{,\mu}$. Similarly, $(\tens{k}\cdot\stcovd \tens{k})^\mu = k^\nu\,{}^\st\nabla_{\!\nu} k^\mu=k^\nu k^\mu{}_{;\nu}$. We use the Greek letters for spacetime indices, and the Latin letters for transverse-space indices. The flat symbol `${\lwix}$\,' 
indicates lowering 
of the tensor indices using the spacetime metric ${\tens{g}}$. Since we will also use the transverse metric ${\mtr}$ for raising and lowering of indices, we denote such spacetime operation explicitly.
}
\begin{equation}\label{kurel}
    \tens{k}\cdot\stgrad u = 0 \;,\quad \lwix\tens{k} = -\, \stgrad u \;.
\end{equation}
The corresponding nontwisting null congruence is automatically geodesic (${\tens{k}\cdot\stcovd \tens{k} =0}$) and we use its affine parameter as another coordinate ${r}$, i.e., we assume
\begin{equation}\label{krrel}
    \tens{k}\cdot \stgrad r =1 \;.
\end{equation}

Intersections of the hypersurfaces ${r=\text{constant}}$ with ${S_u}$ form exactly the above mentioned ${(D{-}2)}$-dimensional transverse spaces, which we denote ${N_{u,r}}$. We assume that these spaces are mutually diffeomorphic, at least in some domain of the spacetime (the situation could be more complicated, e.g., in the presence of various ``black objects''). Therefore, it is natural to identify all the spaces ${N_{u,r}}$ with one \emph{typical transverse space~${N}$}. The space ${N}$ should be understood as a separate Riemannian manifold of dimension ${(D-2)}$ with metric~${\mtr}$ which, for each values of ${u}$ and ${r}$, is embedded into the full Kundt spacetime as ${N_{u,r}}$.

To establish such an embedding explicitly we have to identify the related points in all the transverse spaces ${N_{u,r}}$ which have different values of~${u}$ and~${r}$. For a fixed~${u}$ and different values of ${r}$ it is natural to identify the points along the orbits of the congruence ${\tens{k}}$. For different values of~${u}$, one has to introduce an additional flow in the \mbox{${u}$-direction} which preserves the transverse foliation ${N}$, and also commutes with the flow along the privileged null congruence~${\tens{k}}$. Such flow can be conveniently introduced by a vector field ${\tens{w}}$ tangent to ${r=\text{constant}}$, which satisfies relations
\begin{equation}\label{wcond}
    \tens{w}\cdot \stgrad r =0\;,\quad \tens{w}\cdot\stgrad u = 1\;,\quad [\tens{k},\tens{w}] = 0 \;.
\end{equation}
The two vector fields ${\tens{k}}$ and ${\tens{w}}$ thus span 2-dimensional \emph{temporal surfaces} which form the foliation ${T}$.
Any of these temporal surfaces intersects each spatial transverse space ${N_{u,r}}$ in a single point. We identify  points which belong to the same temporal surface, and map them into one point of the typical transverse space ${N}$. We denote the temporal surface corresponding to a point ${x\in N}$ as~${T_x}$.

The temporal foliation ${T}$ allows us also to identify tensors tangent to the typical transverse space ${N}$ with those spacetime tensors which are trivial on the temporal surfaces. For brevity, we call them \emph{transverse tensors}  (cf.\ Sec.~\ref{sc:trsp}).

The transverse spaces ${N_{u,r}}$ can thus be understood as an embedding of the typical transverse space ${N}$ into the Kundt spacetime. The transverse metric~${\mtr}$ can be viewed as a pullback of the spacetime metric~${\tens{g}}$ into this transverse space. The fact that the congruence~${\tens{k}}$ is nonexpanding and shear-free implies that the transverse part of the metric is conserved along~${\tens{k}}$ (${\lder{\tens{k}} \tens{\mtr}=0}$), which means that ${\mtr}$ is ${r}$-independent.

Explicit identification of the different transverse spaces ${N_{u,r}}$ can be obtained by a convenient choice of the remaining ${(D-2)}$ spatial coordinates ${x^i}$: they can be chosen to be \emph{constant along the temporal surfaces} ${T_{x}}$. With such a natural choice of the adjusted coordinates, the vector fields ${\tens{k}}$ and ${\tens{w}}$ become the coordinate fields
\begin{equation}\label{kwcoor}
    \tens{k}=\cv{r} \;,\quad \tens{w}=\cv{u} \;.
\end{equation}
Also, in these coordinates, the transverse tensors have their ${r}$ and ${u}$ components vanishing.

We may also introduce the \emph{`temporal' derivatives} of a quantity ${\tens{X}}$ along ${\tens{k}}$ and~${\tens{w}}$:
\begin{equation}\label{tempders}
  \rder{\tens{X}} = \lder{\tens{k}} \tens{X}\;,\quad \uder{\tens{X}}=\lder{\tens{w}} \tens{X} \;.
\end{equation}
When restricted to the typical transverse space ${N}$, these turn out to be just derivatives with respect to the parameters ${r}$ and ${u}$, respectively.

Of course, the splitting of the spacetime into the transverse spaces ${N_{u,r}}$ and the temporal surfaces ${T_x}$ is not canonically given by the Kundt geometry. One can choose a different affine coordinate ${r}$ which defines the transverse spaces. Or, one can change the flow ${\tens{w}}$ which identifies the temporal surfaces. These gauge freedoms are shortly discussed in Appendix~\ref{apx:gauge}.

The form of the metric \eqref{metric} indicates that the temporal surfaces ${T_x}$ are not orthogonal to the transverse spaces ${N_{u,r}}$. The non-orthogonality is encoded in the metric 1-form ${\ta}$. Moreover, this property can depend on a particular choice of the gauge. In this paper we assume that the temporal surfaces \emph{could} be chosen to be orthogonal to the transverse spaces,\footnote{%
Geometrically it implies that the transverse spaces ${N_{u,r}}$ can be chosen in such a way that the 2-spaces of vectors orthogonal to ${N_{u,r}}$ are integrable.}
i.e., that the metric 1-form ${\ta}$ could be eliminated by a suitable gauge transformation. Note, however, that we will not use such a privileged gauge choice since otherwise the ${u}$-dependence of the transverse metric ${\mtr}$ would become complicated, see discussion in Sec.~\ref{sc:disc}. Inspecting the gauge behavior \eqref{wgauge} of ${\ta}$ under the transformation \eqref{wshift} we observe that it can be eliminated only provided ${\ta}$ is ${r}$-independent, ${\rder{\ta}=0}$. This is a technical form of our final `integrability' assumption concerning the geometry, mentioned above equation \eqref{metric}.

To summarize, the spacetime metric ${\tens{g}}$ of \eqref{metric} is naturally split into the transverse objects ${H}$, ${\ta}$, and ${\mtr}$ on~${N}$. The remaining assumptions take the explicit form
\begin{equation}\label{assumptions}
  \rder \mtr =0\;,\quad
  \rder \ta =0\;.
\end{equation}
In other words, ${\mtr}$ and ${\ta}$ are ${r}$-independent transverse forms which can both depend on ${u}$~and ${x^i}$. The metric function ${H}$ depends on all the coordinates ${r}$, ${u}$, and~${x^i}$.

\subsection*{Electromagnetic field}

We intend to study the gravitational field generated by null fluid and gyratonic matter in the presence of an aligned uniform electromagnetic field. The `uniformity' requirement will be discussed after the field equations are formulated, cf.\ Secs.~\ref{sc:disc} and \ref{sc:solvable}, specifically equations \eqref{EBis0}, \eqref{EBconst}.

The alignment condition we impose reads that the congruence ${\tens{k}}$ is an eigenvector of the Maxwell tensor ${\tens{F}}$:
\begin{equation}\label{EMaligned}
  \tens{F}\cdot \tens{k} = E\; \lwix \tens{k}\;.
\end{equation}
Consequently, the Maxwell 2-form ${\tens{F}}$ has the form
\begin{equation}\label{Maxwell}
  \tens{F} = E\; \stgrad r\wedge \stgrad u+\stgrad u \wedge \ts + \Bt \;,
\end{equation}
where ${\ts}$ is a transverse \mbox{1-form} and ${\Bt}$ is a transverse \mbox{2-form}. We interpret the first term as an electric part of the field, and ${\Bt}$ as a (transverse) magnetic part, although such an interpretation is not straightforward due to the \mbox{2-dimensional} character of the temporal surfaces ${T_x}$ and ${(D{-}2)}$-dimensional character of the transverse spaces~${N_{u,r}}$.

The stress-energy tensor corresponding to \eqref{Maxwell} has the structure\footnote{\label{TEMsplit}%
In coordinate components we have
$\kap\, T^\EM_{uu}=2 H \rho + \kap\epso {(E \ta {-} \ts)^2}$,
${\kap\, T^\EM_{ur}=\rho}$,
${\kap\, T^\EM_{ui}=\tau a_i + \kap\epso(E\,a^j B_{ji} - E s_i - s^j B_{ji})}$, and
${\kap\, T^\EM_{ij}=\kap\epso\bigl(\frac{1}{2} E^2 q_{ij} + B^2{}_{ij}-\frac12 B^2 q_{ij}\bigr)}$.\\
Here, ${\kap}$ is Einstein's gravitational constant and ${\epso}$ is permittivity of vacuum. Usual choices are the Gaussian one (${\kap=8\pi}$, ${\epso=1/4\pi}$) or SI-like (${\kap=1}$, ${\epso=1}$).}
\begin{equation}\label{TEM}
\begin{split}
  \kap\, \tT^{\EM}&=
     \Bigl(2 H \rho + \kap\epso (E \ta - \ts)^2\Bigr)\,\stgrad u \stgrad u + \rho\, \stgrad u \vee\stgrad r\\
     &\quad+\stgrad u \vee \Bigl(\tau \ta + \kap\epso(E\,\ta\cdot \Bt - E \ts - \ts\cdot \Bt)\Bigr)\\
     &\quad+\kap\epso \Bigl(\frac12 E^2 \mtr + \BBt - \frac12\, B^2 \mtr \Bigr)
     \;,
\end{split}\raisetag{14pt}
\end{equation}
where, e.g., ${\tens{s}\cdot \Bt}$ is a transverse 1-form with components ${s^\nu B_{\nu\mu}}$.
For convenience, we also introduced the scalar quantities ${\rho}$ and ${\tau}$ quadratic in ${E}$ and ${B}$ as
\begin{equation}\label{rhotaudef}
\begin{gathered}
\rho = \frac{\kap\epso}2\bigl(E^2+B^2\bigr)\;,\\
\tau = \frac{\kap\epso}2\bigl(E^2-B^2\bigr)\;,\\
\end{gathered}
\end{equation}
in which the scalar square ${B^2}$ of the transverse magnetic 2-form ${\Bt}$ includes the factor ${1/2}$,
\begin{equation}\label{B2def}
    B^2
    =\frac12\, B_{\mu\kappa}B_{\nu\lambda}\,g^{\mu\nu}g^{\kappa\lambda}
    \;.
\end{equation}
We also introduced the tensorial (matrix) square ${\BBt=-\Bt\cdot\Bt}$ of the 2-form ${\Bt}$ via
\begin{equation}\label{BBdef}
    B^2{}_{\mu\nu}= B_{\mu\kappa}B_{\nu\lambda}\,g^{\kappa\lambda}\;.
\end{equation}
In a dimension ${D\neq4}$, the stress-energy tensor is not tracefree. In fact, its trace is characterized by the quantity ${\tau}$:
\begin{equation}\label{Ttrace}
  \kap\, T^\EM_{\mu\nu}\,g^{\mu\nu}=(D-4)\,\tau\;.
\end{equation}

\subsection*{Gyratonic matter}

As a source of the gravitational field, we also admit a generic \emph{gyratonic matter} aligned with the congruence ${\tens{k}}$. The gyratonic matter is a generalization of a null fluid (pure radiation), allowing also its inner spin \cite{Bonnor:1970b,Griffiths:1972,FrolovFursaev:2005,FrolovIsraelZelnikov:2005,KadlecovaZelnikovKrtousPodolsky:2009}. It is described phenomenologically by the stress-energy tensor
\begin{equation}\label{Tgyr}
  \kap\, \tT^\gyr = j_u\, \stgrad u\stgrad u + \stgrad u \vee \tj\;,
\end{equation}
with the scalar energy density ${j_u}$, and the spinning part given by the transverse 1-form ${\tj}$. Clearly, for ${\tj=0}$ we obtain standard null fluid moving along the null direction~${\tens{k}}$.

We do not specify the field equation of the gyratonic matter, except that we assume its local stress-energy conservation
\begin{equation}\label{Tgyrcons}
  \stdiv \tT^\gyr = 0\;.
\end{equation}
In indices this reads ${g^{\mu\nu}\,{}^\st\nabla{}_\mu T^\gyr{}_{\mspace{-10mu}\nu\kappa}=0}$.

\section{Transverse-space formulation}\label{sc:trsp}


\subsection*{Transverse tensors}

In the next section, we will formulate the field equations purely in terms of quantities on the transverse space ${N}$. We have already mentioned that the transverse tensors can be viewed in two closely related ways.

Naturally, these are the quantities from tangent tensor space of the typical transverse space ${N}$, which may depend on two additional parameters ${u}$ and ${r}$.

Alternatively, they can be understood as \emph{spacetime tensors} which are tangent to the embedded transverse spaces ${N_{u,r}}$ and which vanish in directions of the temporal surfaces ${T_x}$. They can be defined using the projector~${\tens{p}}$ onto the transverse space
\begin{equation}\label{transproj}
  \tens{p} = {}^\st\!\tens{\delta}-\tens{k}\,\stgrad r - \tens{w} \, \stgrad u\;,
\end{equation}
where ${{}^\st\!\tens{\delta}}$ is the identity spacetime tensor with components ${\delta^\mu_\nu}$.
This projector leaves unchanged the vectors tangent to ${N_{u,r}}$ and also the 1-forms which
annihilate vectors tangent to the temporal surfaces ${T_x}$.
Equivalently, it annihilates the vectors ${\tens{k}}$ and ${\tens{w}}$ spanning ${T_x}$, and the 1-forms ${\stgrad u}$ and ${\stgrad r}$.
In the adjusted coordinates ${\{u,r,x^i\}}$ such a projection simply cancels all the ${u}$ and ${r}$
tensor components, while leaving the transverse components unchanged.

In the former approach we use the Latin tensor indices.
In the spacetime picture we use the Greek tensor indices even for the transverse tensors because in generic coordinates all components could be nontrivial. Only in the adjusted coordinates  all the ${u}$ and ${r}$ components of the transverse tensors vanish. Therefore, in such coordinates, the expressions containing just the transverse tensors can be easily transformed to the corresponding expressions on ${N}$ just by switching from the Greek indices to the Latin ones. For example, for the transverse 2-form ${\Bt}$ its scalar square \eqref{B2def} can be written using only the transverse indices as ${B^2=\frac12 B_{ik}B_{jl}q^{ij}q^{kl}}$, while the tensorial square \eqref{BBdef} becomes ${B^2{}_{ij}= B_{ik}B_{jl}q^{kl}}$.

The spaces of transverse tensors understood as quantities in full spacetime depend, in general, on a particular choice of gauge. Contrary, the tangent space of the typical transverse manifold ${N}$ is gauge independent. Clearly, the identification of these two pictures (induced by the embedding ${N\to N_{u,r}}$) is gauge-dependent. This dichotomy is the reason why it is useful to keep both these views of the transverse objects. The field equations will be naturally expressed in the language of quantities on the typical transverse space~${N}$. On the other hand, the splitting of spacetime objects and properties of gauge transformations are easier to study employing the spacetime picture.

To reduce a general spacetime tensor to the transverse space ${N}$, first we have to split it  into its temporal and transverse parts. For that, we construct all its temporal projections on the vectors ${\tens{k}}$, ${\tens{w}}$ and the 1-forms ${\stgrad u}$, ${\stgrad r}$, as well as the transverse projections using the projector~${\tens{p}}$. In the notation without components we indicate each transverse projection using the symbol `${\trpr}$' at the position of the projected index. For example, for a vector ${\tens{v}}$ we define ${\tens{v}^\trpr = \tens{p}\cdot \tens{v}}$.

We have already encountered such a splitting of the basic geometric, electromagnetic, and gyratonic quantities in Sec.~\ref{sc:ansatz}. The metric ${\tens{g}}$ splits into the scalar ${uu}$ component ${-2H=g_{uu}}$, the \mbox{${\tens{p}}$-projection} of the ${u}$ component ${\ta=\tens{g}_{u\trpr}}$, and to the transverse metric ${\mtr=\tens{g}_{\trpr\trpr}}$. Similarly, the Maxwell tensor ${\tens{F}}$ splits into ${E=F_{ru}}$, ${\ts=\tens{F}_{u\trpr}}$, and ${\Bt=\tens{F}_{\trpr\trpr}}$, with the transverse tensors ${\ts}$ and ${\Bt}$. The splitting of the gyratonic stress-energy tensor gives ${j_u = T^\gyr_{uu}}$ and
${\tj=\tT^\gyr_{u\trpr}}$. The splitting of the electromagnetic stress-energy tensor can be read out from expression \eqref{TEM}, or, in components, it is given in footnote \ref{TEMsplit}.

We must also study the relation between spacetime and transverse-space derivatives. We already introduced the temporal derivatives \eqref{tempders} along ${\tens{k}}$ and ${\tens{w}}$. In the spacetime they correspond to the Lie derivatives, in the transverse-space formulation they are just parametric derivatives with respect to ${r}$ and ${u}$. The spacetime gradient of a scalar function ${f}$ can thus be split into its temporal and transverse parts
\begin{equation}\label{stgradsplit}
    \stgrad f = \rder f\, \stgrad r + \uder f\,\stgrad u+\grad f\;.
\end{equation}
The transverse-space gradient ${\grad f}$ is the ${\tens{p}}$-projection of the spacetime gradient,
\begin{equation}\label{stgradproj}
    \grad f = \tens{p}\cdot \stgrad f 
    \;.
\end{equation}
The same relation holds for the exterior derivative of transverse antisymmetric forms \cite{KrtousPodolsky:inprep}.

Moreover, under the condition ${\rder \mtr =0}$, the transverse-space covariant derivative ${\covd \tens{A}}$ of a transverse tensor ${\tens{A}}$ is also given by the ${\tens{p}}$-projection of the spacetime derivative ${\stcovd \tens{A}}$,
\begin{equation}\label{covdproj}
    \covd \tens{A}
    = \bigl(\stcovd^{} \tens{A}\bigr){}^{\;\,\trpr\dots}_{\trpr\trpr\dots}\;.
\end{equation}
This can be checked, e.g., in adjusted coordinates by inspecting the Christoffel symbols involved \cite{PodolskyZofka:2009,KrtousPodolsky:inprep}.

\subsection*{Splitting of the curvature}

To express the Einstein equations in terms of quantities on the typical transverse space ${N}$ we need to find the projections of the spacetime Ricci tensor ${\stRic}$. For a general Kundt class,
they have been explicitly calculated in components in \cite{PodolskyZofka:2009} and
expressed in the covariant form in \cite{KrtousPodolsky:inprep}.
Assuming \eqref{assumptions}, different projections restricted on the transverse space are
\begin{align}
  \stRic_{rr}&=0\;,\notag\\
  \stRic_{r\trpr} &= 0\;,\notag\\
  \stRic_{ru} &= {\rdder{H}}\;,\label{Riccisplit}\\
  \stRic_{uu} &=
    \LB H
    + f^2
    +2 {\rdder H}\Bigl(H{+}\frac12 a^2\Bigr) \notag\\
    &\quad+ \rder H \div \ta + 2 \ta \cdot \grad \rder H - \rder H \Theta \notag\\
    &\quad+\div {\uder \ta}
    - {\uder q}^2
    - {\uder \Theta} \;,\notag\\
  \stRic_{u\trpr} &= -\frac12 \div\tf + \grad \rder H
     +\frac12 \div{\uder \mtr} - \grad \Theta \;,\notag\\
  \stRic_{\trpr\trpr} &= \Ric\;.\notag
\end{align}
The spacetime scalar curvature ${\stscR}$, expressed in terms of the transverse scalar curvature ${\scR}$, is
\begin{equation}\label{scRsplit}
    \stscR = -2\rdder H+ \scR\;.
\end{equation}
Here ${\Ric}$ and ${\scR}$ are the Ricci tensor and scalar curvature of the transverse metric ${\mtr}$, respectively.
We also introduced the abbreviations
\begin{gather}\label{fdef}
    \tf=\grad\ta\;,\\
    a^2 = \ta\bullet\ta = \ta\cdot\ta= a_i\, a^i\;,\\
    f^2 = \tf\bullet\tf = \frac12 f_{ij} f^{ij}\;,\\
    \Theta = \mtr\bullet \uder\mtr = \frac12\,  q^{ij}\,{\uder q}_{ij}  = \ivol(\vol)\,\uder{}\;.
    \label{Thetadef}
\end{gather}
Clearly, ${\Theta}$ characterizes the rate of ${u}$-change of the transverse volume element ${\vol=(\Det \mtr)^{\frac12}}$.

\subsection*{Useful identities on the transverse space}

In the above equations we have employed the form product ${\bullet}$, the transverse Laplace--Beltrami operator ${\LB}$, and the transverse divergence ${\div}$.
In this section we briefly review some related definitions and identities.

Recall that the typical transverse space ${N}$ is a ${d}$-dimensional Riemannian space with a metric ${\mtr}$, where ${d=D{-}2}$. We use this metric to lower and raise the Latin indices and we do this operation without any indication.\footnote{%
Essentially, we do not distinguish between the transverse forms and vectors. Since the metric ${\mtr}$ is non-degenerate, it does not usually lead to any confusion.
One has to be careful only in situations when the metric ${\mtr}$ changes with the external parameter ${u}$. In such a case lowering and rising of indices does not commute with the ${u}$-derivative.}

The metric ${\mtr}$ and a chosen orientation fixes the \emph{Levi-Civita tensor} ${\LC}$ which allows us to define the \emph{Hodge dual} of an antisymmetric ${p}$-form ${\tens{\omega}}$:
\begin{equation}\label{hodge}
  (*\omega)_{a_{p+1}\dots a_d}=\frac1{p!}\,\omega^{a_1\dots a_p}\LCC_{a_1\dots a_d}\;.
\end{equation}
It satisfies
\begin{equation}\label{invhodge}
  * * \tens{\omega} = (-1)^{p(d-p)}\tens{\omega}\;,
\end{equation}
where we assume the positive definiteness of ${\mtr}$. Consequently, ${*^{\!-\!1}\tens{\omega}=(-1)^{p(d-p)}*\tens{\omega}}$.

The \emph{inner product} on antisymmetric ${p}$-forms is defined
\begin{equation}\label{bulletdef}
  \tens{\omega}\bullet\tens{\sigma} = \frac1{p!}\,\omega_{a_1\dots a_p} \sigma^{a_1\dots a_p}\;.
\end{equation}
This satisfies the relation $\tens{\omega}\wedge(*\tens{\sigma})=\tens{\sigma}\wedge(*\tens{\omega})=(\tens{\omega}\bullet\tens{\sigma})\,\LC$.
We will use the definition \eqref{bulletdef} also for symmetric ${p}$-forms, such as in \eqref{Thetadef}.

We employ an ordinary dot `${\cdot}$' symbol to indicate a contraction in just one index. For example a scalar ${a^2}$ is given by ${\ta\cdot\ta = a_i\, a^i}$, 1-form  ${\ta\cdot\Bt}$ has components ${a^i B_{ij}}$
(but ${\Bt\cdot\ta}$ has components ${B_{ij} a^j}$), and components of 2-form ${\Bt\cdot\Bt}$ are ${B_{ik} B^{k}{}_{j}}$, cf.~\eqref{BBdef}.

We define the \emph{transverse divergence} of a general ${p}$-form
\begin{equation}\label{divdefix}
  (\div \omega)_{a_1\dots a_{p-1}} = \nabla{}^i \omega_{ia_1\dots a_{p-1}}\;.
\end{equation}
For antisymmetric ${p}$-forms the divergence is, up to a sign, the standard co-derivative ${\delta}$:
\begin{equation}\label{divdelta}
  \div\tens{\omega} = \covd\cdot\,\tens{\omega} = -\delta\tens{\omega}=-(-1)^{p}*^{\!-\!1}\!\grad\!*\tens{\omega}\;.
\end{equation}
Consequently, ${\div\div\tens{\omega}=0}$.

We define the \emph{Laplace--de~Rham operator} on antisymmetric forms as\footnote{%
In our convention ${\lapl}$ is a negative-definite operator and it has the same sign as the Laplace--Beltrami operator, cf.\ eq.~\eqref{WBident}.}
\begin{equation}\label{LRdef}
  \lapl=\grad \div+\div\grad\;.
\end{equation}
This is related to the \emph{Laplace--Beltrami operator}
\begin{equation}\label{LBdef}
  \LB=\covd\cdot\covd = q^{ij}\nabla_i\nabla_j
\end{equation}
through the Weitzenb\"ock--Bochner identity
\begin{equation}\label{WBident}
\begin{split}
  \lapl\omega_{a_1\dots a_p}&= \LB\omega_{a_1\dots a_p}
    -p\,\RicC{}_{n[a_1}\omega^n{}_{a_2\dots a_p]}\\
    &\qquad+{\textstyle\frac{p(p{-}1)}2}\; R_{mn[a_1a_2}\omega^{mn}{}_{a_3\dots a_p]}\;.
\end{split}
\end{equation}
In particular, for a scalar ${h}$, there is ${\lapl h = \LB h}$.

The Hodge theory tells us that any form ${\tens{\omega}}$ can be written using its \emph{potential} ${\tens{\alpha}}$, \emph{co-potential} ${\tens{\beta}}$, and \emph{harmonic} ${\tens{\omega}_\harm}$ as
\begin{equation}\label{HodgeTh}
    \tens{\omega} = \grad \tens{\alpha} + \div\tens{\beta} + \tens{\omega}_\harm\;.
\end{equation}
The potentials can be restricted by additional gauge conditions
\begin{equation}\label{HTgauge}
    \div\tens{\alpha}=0\;,\quad\grad\tens{\beta}=0\;.
\end{equation}
The harmonic part satisfies
\begin{equation}\label{harm}
    \grad \tens{\omega}_\harm =0\;,\quad \div\tens{\omega}_\harm =0\;,
\end{equation}
which, of course, implies ${\lapl\tens{\omega}_\harm=0}$.

For compact spaces the splitting \eqref{HodgeTh} is unique and \eqref{harm} is equivalent to ${\lapl\tens{\omega}_\harm=0}$. For noncompact spaces this splitting is not unique and \eqref{harm} is a stronger restriction then ${\lapl\tens{\omega}_\harm=0}$. However, even in the noncompact cases, reasonable boundary conditions can guarantee uniqueness of the Hodge splitting. Usually, we will assume such conditions to be satisfied, at least for the background gravitational and electromagnetic fields.

\section{The Field equations}\label{sc:fieldeq}

In this section we present the field equations in the transverse-space formalism.
We also explicitly solve the \mbox{${r}$-dependence} of the fields.

\subsection*{The Maxwell equations}

Transverse projections of various components of the Maxwell equations
\begin{equation}\label{stMaxwell}
 \stgrad \tens{F} =0 \;,\quad  \stdiv \tens{F} = 0
\end{equation}
give the equations \cite{PodolskyZofka:2009,KrtousPodolsky:inprep}
\begin{equation}\label{trMaxwell1}
\begin{gathered}
  \rder \Bt =0\;,\quad \grad \Bt = 0\;,\\
  \rder \ts = - \grad E\;,\quad  \grad \ts = \uder \Bt \;,
\end{gathered}
\end{equation}
and
\begin{equation}\label{trMaxwell2}
\begin{gathered}
  \rder E =0\;,\\
  \bigl(E \ta +\ta\cdot \Bt -\ts\bigr)\,\rder{} = - \div \Bt \;,\\
  \div\bigl(E \ta +\ta\cdot \Bt -\ts\bigr) = \uder E + \Theta E\;,
\end{gathered}
\end{equation}
respectively. We solve the ${r}$-dependence of ${\ts}$ by setting
\begin{equation}\label{sigmadef}
  \ts= -r\,\grad E + \tsg\;,\quad\text{where}\quad\rder\tsg=0\;.
\end{equation}
The Maxwell equations are then equivalent to
\begin{subequations}\label{trMaxwell}
\begin{gather}
  \rder E =0 \;,\quad \rder \Bt =0\;,\quad \rder\tsg = 0\;,\label{MEr}\\
  \uder \Bt = \grad \tsg\;,\quad\grad \Bt = 0\;,\label{MEB}\\
  \grad E + \div \Bt = 0\;,\label{MEEB}\\
  \div\bigl(E \ta +\ta\cdot \Bt -\tsg\bigr) = \uder E + \Theta E\;,\label{MEEBs}
\end{gather}
\end{subequations}
where we used ${\div(\div \Bt)=0}$.

Let us note that, as a consequence of the Maxwell equations, we may also split the 1-form ${\grad E\cdot \Bt}$ into its gradient and divergence parts:
\begin{equation}\label{dEB}
\grad E \cdot  \Bt = \div(E \Bt)+\frac12\grad(E^2)\;.
\end{equation}

\subsection*{Gyraton stress-energy conservation}

The divergence of the stress-energy tensor \eqref{Tgyr} can be written as
\begin{equation}\label{divT}
  \kap \stdiv \tT^\gyr = \bigl((-j_u+\tj\cdot\ta)\,\rder{} + \div \tj\bigr) \stgrad u - (\tj)\,\rder{}\;.
\end{equation}
The condition \eqref{Tgyrcons} can thus be solved by setting
\begin{equation}\label{iotadef}
  j_u = r\div \tj + \iota
\end{equation}
with
\begin{equation}\label{jiotacond}
  (\iota)\,\rder{}= 0 \;,\quad (\tj)\,\rder{} =0\;.
\end{equation}

\subsection*{The Einstein equations}

Finally, we split the Einstein equations with the cosmological constant ${\Lambda}$
\begin{equation}\label{Einstein}
  \stRic -\frac12\stscR\, \tens{g}+\Lambda\, \tens{g} = \kap\, \tT\;,
\end{equation}
where the total stress-energy tensor  ${\tT = \tT^\EM+\tT^\gyr}$ has the form
\begin{equation}\label{Ttot}
  \kap\, \tT =  \rho^\tot\, \stgrad u\vee\stgrad r + j_u^\tot\, \stgrad u \stgrad u + \stgrad u \vee \tj^\tot + \kap\, \tT_{\trpr\trpr}\;,
\end{equation}
with the transverse components given by
\begin{align}
  \rho^\tot &= \rho\;,\quad \label{Tsplitrho}\\
  j_u^\tot &= 2H \rho + \kap\epso (E\ta-\ts)^2 + r \div \tj + \iota\;,\label{Tsplitju}\\
  \tj^\tot &= -\rho \ta - \kap\epso (\ts - E\ta)\cdot(E \mtr + \Bt)+ \tj\;,\label{Tsplitj}\\
  \tT_{\trpr\trpr} &= 
  \frac\epso2 E^2 \mtr + \epso\bigl(\BBt-\frac12B^2\mtr\bigr) \;.\label{TsplitT}
\end{align}
Since the gyratonic stress-energy tensor is trace-free, we have
\begin{equation}\label{taudef}
  (D-4)\tau = \kap\, T^\mu_\mu \;.
\end{equation}
Trace and trace-free parts of the total transverse stress-energy tensor \eqref{TsplitT} are
\begin{gather}
  \kap\, T_{\trpr\trpr}{}^i_i = 2\rho + (D-4)\, \tau\;,\label{traceT}\\
  \frac1\epso \tT_{\trpr\trpr}^{\,\TF} = \BBt - \frac{2}{D{-}2}\, B^2 \mtr\;.\label{tracefreeT}
\end{gather}

Now, we will perform the first steps in integration of the Einstein equations.

\subsubsection*{Metric function ${H}$}

Substituting \eqref{Riccisplit}, \eqref{scRsplit} and \eqref{Ttot} into \eqref{Einstein}, we easily check that the components ${rr}$ and ${r\trpr}$ of the Einstein equations are trivially satisfied. The ${ru}$ component gives
\begin{equation}\label{EEru}
  \frac12\scR = \rho + \Lambda\;.
\end{equation}
The trace of the Einstein equations implies
\begin{equation}\label{EEtrace}
  \frac12\scR-\rdder H = -\frac{D-4}{D-2}\tau + \frac{D}{D-2}\Lambda\;.
\end{equation}
Eliminating the transverse scalar curvature ${\scR}$ from these two equations we obtain the equation for ${\rdder H}$:
\begin{equation}\label{hrdder}
    \rdder H =  \rho + \frac{D-4}{D-2}\,\tau -\frac2{D-2}\, \Lambda \;.
\end{equation}
Taking into account the first two of the equations \eqref{trMaxwell} we find that ${H}$ can be integrated explicitly to
\begin{equation}\label{rdependofH}
    H = \frac12\Bigl(\rho + \frac{D-4}{D-2}\,\tau-\frac2{D-2}\, \Lambda\Bigr)\, r^2 + g\, r + h\;,
\end{equation}
where ${\rho}$ and ${\tau}$ are defined in \eqref{rhotaudef}, and the functions ${g}$ and ${h}$ are (possibly ${u}$-dependent) scalar functions on the transverse space ${N}$.

\subsubsection*{Equation for the transverse metric ${\mtr}$}

The trace of the transverse part of the Einstein equations is a linear combination of equations \eqref{EEru} and \eqref{EEtrace}. The trace-free part together with equation \eqref{EEru} give the following equation for the transverse metric:
\begin{equation}\label{EEq}
    \Ric = \frac{2}{D-2}(\rho+\Lambda)\,\mtr + \kap\, \tT_{\trpr\trpr}^{\,\TF}\;,
\end{equation}
where ${\tT_{\trpr\trpr}^{\,\TF}}$ is determined by \eqref{tracefreeT}.
As a consequence of vanishing divergence of the Einstein tensor one also obtains
\begin{equation}\label{divEB}
    \div(E\Bt) = \frac1{D{-}2}\,\grad\tau\;.
\end{equation}

\subsubsection*{Equation for the metric 1-form ${\ta}$}

The ${u\trpr}$ component of Einstein's equations gives:
\begin{equation}\label{EEuT}
\begin{split}
    &-\frac12 \div \tf  + \rdder H \ta
    +\kap\epso (\ts -E \ta) \cdot (E \mtr+ \Bt) + \grad\rder H\\
    &\qquad\qquad
    =\grad \Theta - \frac12 \div \uder \mtr + \tj\;.
\end{split}\raisetag{18pt}
\end{equation}
This expression is linear in $r$ (with the $r$-dependence hidden just in $\rder H$ and $\ts$).
Using \eqref{divEB} it can be shown that the $r$ term is a consequence of the already known field equations. The $r$-independent part gives the equation for~$\ta$:
\begin{equation}\label{aeq}
\begin{split}
    &-\frac12 \div \tf  + \rdder H \ta
    +\kap\epso (\tsg -E \ta) \cdot (E \mtr+ \Bt) + \grad g\\
    &\qquad\qquad
    =\grad \Theta - \frac12 \div \uder \mtr + \tj\;.
\end{split}\raisetag{18pt}
\end{equation}

\subsubsection*{Equation for ${g}$}

Finally, the ${uu}$ component leads to an expression quadratic in $r$. Using \eqref{dEB}, \eqref{divEB}, and the fact that $E$ is harmonic (which is a consequence of \eqref{MEEB}) it is possible to show that the quadratic term is trivial. The linear term in ${r}$ gives:
\begin{equation}\label{laplg}
\begin{split}
    &\lapl g + 2 \ta\cdot\grad\rdder H + (\div \ta - \Theta)\rdder H \\
    &\qquad\qquad+2\kap\epso (\tsg -E \ta) \cdot \grad E = \div \tj\;.
\end{split}
\end{equation}
In fact, it turns out that this equation is equivalent to the divergence of \eqref{aeq}. To show this, one has to employ the trace of the $u$-derivative of \eqref{EEq} and the geometrical relation
\begin{equation}\label{ricuder}
  {\uder\RicC}_{ij}\;q^{ij} = -2\lapl\Theta + \frac12\nabla^{\,i}\nabla^{\,j}\uder{q}_{ij}\;.
\end{equation}
which follows from the fact that, thanks to relation ${\covd \mtr =0}$, the transverse covariant derivative can be $u$-dependent.

\subsubsection*{Equation for ${h}$}

Finally, the remaining $r$-independent part of  the ${uu}$ component of the Einstein equations gives the equation for $h$:
\begin{equation}\label{heq}
\begin{split}
    &\lapl h + 2 \ta\cdot\grad g + \rdder H a^2 + (\div \ta {-} \Theta)\,g \\
    &\qquad - \kap\epso (\tsg {-}E \ta)^2 +\frac12 f^2 = \frac12 {\uder q}^2 - \div \uder{\ta} + \uder \Theta + \iota\;.
\end{split}\raisetag{8ex}
\end{equation}

\section{Discussion of the equations}\label{sc:disc}

\subsection*{Decoupling of the equations and adjusting the electromagnetic field to the geometry}

The Einstein equations for the transverse metric \eqref{EEq} and the Maxwell equations for the electromagnetic field \eqref{trMaxwell} are coupled, so that they cannot be solved one after another. This significantly complicates the process of finding explicit solutions. However, we can restrict the generality of the electromagnetic field in such a way that a solvable system is obtained, which describes the evolution of a gyratonic matter accompanied by the gravitational wave in a non-dynamical electromagnetic field.

Namely, we will restrict ourselves to the cases when the right-hand side of \eqref{EEq} is given just by tensors obtained in an algebraic way from the transverse metric ${\mtr}$. For this we have to assume some special geometrical and/or topological structure of the transverse space ${N}$, and suitably adjusted uniform electromagnetic field.

In Sec.~\ref{sc:solvable}, we will discuss two important explicit examples. In both of them the electric field is taken to be constant, ${E=\text{constant}}$. In the first case we will assume that the magnetic part is completely missing, ${\Bt=0}$. In the second case we will assume that the geometry of the transverse space ${N}$ is a direct product of 2-dimensional spaces, and the magnetic field is given by a linear combination of canonic 2-forms on these 2-dimensional components.

In both these cases the right-hand side of \eqref{EEq} only depends on a finite number of constants characterizing the electromagnetic field and on the preselected form of the transverse geometry. Choosing these electromagnetic constants we can find the corresponding transverse geometry, and on this background to solve the remaining field equations.

Actually, such restrictions imposed on the electromagnetic field are not excessively strong. Taking the gradient and divergence of equations \eqref{MEEB} together with \eqref{MEB} we find that both ${E}$ and ${\Bt}$ must be a harmonic 0-form and a 2-form, respectively,
\begin{equation}\label{harmEB}
    \lapl E=0\;,\quad \lapl \Bt =0 \;.
\end{equation}
Imposing the natural assumption of finiteness of the fields at infinity of the transverse space, or restricting to compact transverse spaces, mathematical theorems guarantees that ${E}$ is constant and that ${\Bt}$ can be nontrivial only for some topologically special spaces. If we, in  addition, assume that both ${\rho}$ and ${\tau}$ are constants (which simplifies the structure of the function ${H}$, cf.\ eq.~\eqref{hrdder}), we obtain the condition ${B^2=\text{constant}}$. Nontrivial harmonic \mbox{2-forms}~${\Bt}$ with a constant square can only exist in very special spaces, of which the direct-product spaces discussed below are significant representatives.

\subsection*{Backgrounds}

To build up a physical intuition for more complicated solutions, it is convenient first to distinguish an electrovacuum non-dynamical \emph{background geometry} characterized by the transverse metric ${\mtr}$, the electric scalar ${E}$, and the magnetic 2-form ${\Bt}$. Moreover, we assume that these electromagnetic quantities are `uniform' in the sense that they are characterized just by a finite number of constant parameters.

Subsequently, the dynamical geometries will be obtained as a `deformation' of these backgrounds due to the presence of gravitational waves and gyratonic matter. Such dynamical degrees of freedom will be described by the metric 1-form ${\ta}$, the metric scalars ${g}$ and ${h}$ (introduced in \eqref{rdependofH}), and the electromagnetic 1-form~${\tsg}$ (introduced in \eqref{sigmadef}).

Specifically, by the background geometry we understand the metric \eqref{metric} with
\begin{equation}\label{backgroundg}
    \ta=0\;,\quad g=0\;,\quad h=0\;,
\end{equation}
accompanied by the electromagnetic field \eqref{Maxwell} with
\begin{equation}\label{backgroundEM}
    \tsg=0
\end{equation}
and vanishing gyratonic matter,
\begin{equation}\label{bacgroundgyr}
    j_u=0\;,\quad \tj=0\;.
\end{equation}
The field equations simplify considerably if ${\rho}$ and ${\tau}$ are transverse-space constants. In view of \eqref{rhotaudef} we thus assume
\begin{equation}\label{ptconst}
    E=\text{constant}\;,\quad
    B^2=\text{constant}\;.
\end{equation}
This implies that ${E}$ and ${\Bt}$ have the harmonic character,
\begin{equation}\label{EBharm}
  \grad E=0\;,\quad\grad \Bt = 0\;,\quad\div\Bt=0\;,
\end{equation}
see relations \eqref{trMaxwell} and \eqref{harm}.\footnote{%
In the compact cases, or assuming sufficiently strong boundary conditions in the noncompact case, equation \eqref{MEEB} already enforces \eqref{EBharm} without a~priori assuming that ${E}$ is constant.}
The transverse metric ${\mtr}$ must then satisfy equations \eqref{EEq}, \eqref{tracefreeT}:
\begin{equation}\label{qequation}
    \Ric = \frac{2}{D{-}2}(\rho+\Lambda)\,\mtr +
    \kap\epso \Bigl( \BBt - \frac{2}{D{-}2}\, B^2 \mtr\Bigr)\;.
\end{equation}

Finally, we have to deal with a possible ${u}$-dependence of the background quantities. To simplify the analysis of the field equations we restrict ourselves  to the simplest case
\begin{equation}\label{uind}
    \uder\mtr=0\;,\quad \uder E = 0 \;,\quad \uder \Bt=0\;.
\end{equation}
The expansion parameter \eqref{Thetadef} then vanishes, ${\Theta=0}$. Notice that in four spacetime dimensions the assumption ${\uder\mtr=0}$ is not necessary: this could always be locally achieved by a suitable gauge transformation. In higher dimensions, this is generally not possible.

To summarize, the metric of the background spacetimes reads
\begin{equation}\label{metricbkgr}
  \tens{g} = \Lambda_- r^2\, \stgrad u\, \stgrad u - \stgrad u \vee \stgrad r + \mtr\;,
\end{equation}
where the constant ${\Lambda_-}$ is
\begin{equation}\label{Lambda-}
    \Lambda_- = \frac2{D-2}\, \Lambda - \rho - \frac{D-4}{D-2}\,\tau\;,
\end{equation}
(so that ${\rdder H = - \Lambda_-}$, see \eqref{hrdder})
and the Maxwell tensor is
\begin{equation}\label{Maxwellbkgr}
  \tens{F} = E\; \stgrad r\wedge \stgrad u + \Bt \;.
\end{equation}

The spacetimes \eqref{metricbkgr}--\eqref{Maxwellbkgr} which satisfy \eqref{qequation} are of \emph{type D} or are \emph{conformally flat}. Indeed, the only nonvanishing components of the Weyl tensor in the natural null frame ${\tens{m}_0=\tens{k}=\cv{r}}$, ${\tens{m}_1=-\frac12\Lambda_- r^2\,\cv{r}-\cv{u}}$, ${\tens{m}_i=m_i^k\,\cv{k}}$ (where the transverse vectors are normalized as ${\,q_{kl}\,m_i^k m_j^l=\delta_{ij}\,}$) are
\begin{eqnarray}
  {}^\st C_{0101}\!\!&=&\!\! \frac{1}{D-1}\!\left[\frac{D-4}{D-2}\big((D-1)\rho+(D-3)\tau  \big)-2\Lambda\right]\!,\nonumber\\
  {}^\st C_{0i1j}\!\!&=&\!\!  \frac{1}{D-2}\big({}^\st C_{0101}\,\delta_{ij}-\kap\, T_{\trpr\trpr}^{\,\TF} {}_{ij}\big)\,,\label{Weylframecomp}\\
  {}^\st C_{ijkl}\!\!&=&\!\! C\,{}_{ijkl}\,,\nonumber
\end{eqnarray}
where the trace-free part ${ \tT_{\trpr\trpr}^{\,\TF} }$ of the transverse stress-energy tensor is defined in \eqref{tracefreeT}, and ${\tens{C}}$ is the Weyl tensor of the  transverse metric ${\mtr}$. All these Weyl scalars are of boost weight 2 and thus the spacetimes are of type~D. In the particular case when all of them vanish the spacetimes are conformally flat. This occurs if, and only if, the cosmological constant is uniquely related to the uniform electric and magnetic fields as ${\Lambda=\frac{1}{2}\kap\epso(D-4)\bigl(E^2+\frac{1}{D-2}B^2\bigr)}$, the trace-free part of the total transverse stress-energy tensor vanishes (${ \tT_{\trpr\trpr}^{\,\TF} =0}$), and the transverse space is conformally flat (${\tens{C}=0}$).

\subsection*{The field equations for gravitational waves and gyratons}

`Non-background' solutions with gravitational waves and gyratons which we will consider do not change the background quantities ${\mtr}$, ${E}$, and ${\Bt}$. However, such solutions will have a different spacetime geometry from the background, since the metric \eqref{metric} will additionally contain the metric 1-form ${\ta}$, the scalars ${g}$ and ${h}$, and the electromagnetic field \eqref{Maxwell} will contain the term with 1-form ${\tsg}$.

The remaining nontrivial Maxwell equations \eqref{MEB} and \eqref{MEEBs} reduce to
\begin{gather}
  \grad\tsg=0\;,\label{MEsui}\\
  \div\bigl(E \ta +\ta\cdot \Bt -\tsg\bigr) = 0\;.\label{MEEBsui}
\end{gather}
The equations \eqref{aeq}, \eqref{laplg}, and \eqref{heq} for ${\ta}$, ${g}$, and ${h}$ simplify to
\begin{equation}\label{aequi}
    \frac12 \div \tf  + \Lambda_- \ta
    -\kap\epso (\tsg -E \ta) \cdot (E \mtr+ \Bt) - \grad g
    =- \tj\;,
\end{equation}
\begin{equation}\label{laplgui}
    \lapl g - \Lambda_- \div \ta = \div \tj\;,
\end{equation}
and
\begin{equation}\label{hequi}
    \lapl h = \iota - 2 \ta\cdot\grad g +\Lambda_- a^2 - g \div \ta
    + \kap\epso (\tsg {-}E \ta)^2 -\frac12 f^2 \;,
\end{equation}
respectively.

Now, we can distinguish two important classes of such solutions. The first is the class of pure gravitational waves, propagating on the backgrounds \eqref{metricbkgr}, characterized by the absence of the gyratonic-matter source, ${\tj=0}$.

The second, more general class of solutions, represents the gravitational and electromagnetic response to the nontrivial gyratonic matter ${\tj}$. These solutions thus describe external fields around the beam of null fluid with inner spin.

\subsection*{Geometry of the temporal surfaces ${T_x}$}

Until now, we have concentrated mostly on the geometry of the Riemannian transverse space ${N}$, because it was convenient to formulate the field equations in terms of quantities on this space. However, it is also important to investigate the geometry of the complementary temporal surface ${T_x}$. The restriction of the spacetime metric \eqref{metric} to these surfaces yields the Lorentzian metric
\begin{equation}\label{tempmtr}
    \tens{g}|_{T_x} = - 2H\, \stgrad u\stgrad u - \stgrad u \vee \stgrad r\;.
\end{equation}
The only nontrivial curvature characteristic of any \mbox{2-dimensional} metric is the Gaussian curvature~${K}$. For the metric \eqref{tempmtr} it turns to be ${K= - \rdder H}$. Substituting the definitions \eqref{rdependofH} and \eqref{Lambda-}, we obtain
\begin{equation}\label{GaussTemp}
    K = \Lambda_-\;.
\end{equation}

It is interesting that the temporal geometry only depends on the single background parameter ${\Lambda_-}$, i.e., it remains the same even if the gravitational wave or gyratonic contributions are present.

In particular, we find that the background spacetimes have the direct-product form ${M=T\times N}$, with the geometry \eqref{metricbkgr} given as the product of the 2-dimensional temporal component ${T}$ of constant Gaussian curvature ${\Lambda_-}$, and the transverse component ${N}$ with the metric ${\mtr}$ satisfying \eqref{qequation}.

For such backgrounds, it is possible to perform the transformation of the temporal metric \eqref{tempmtr} to the canonical form. Namely, the transformation
\begin{equation}\label{UVtrans}
    U = \frac1{\Lambda_- u}\;,\quad V = \frac4{\Lambda_- r} + 2u\;,
\end{equation}
(or ${U=u}$ and ${V=r}$ for ${\Lambda_-=0}$) leads to
\begin{equation}\label{tempmtrcan}
    \tens{g}|_{T_x} = -\bigl(1-{\textstyle\frac12}\Lambda_-\,UV\bigr)^{\!-2}\;\stgrad U\vee\stgrad V\;.
\end{equation}
According to the sign of the constant ${\Lambda_-}$ there are three possibilities: for ${\Lambda_-=0}$ the temporal surface ${T_x}$ is the 2-dimensional Minkowski space ${M_2}$, for ${\Lambda_->0}$ it is the \mbox{2-dimensional} de~Sitter space ${dS_2}$, and for ${\Lambda_-<0}$ we get the 2-dimensional anti-de~Sitter space ${AdS_2}$. The characteristic scale of the (anti-)de~Sitter space is given by ${\ell=1/\sqrt{|\Lambda_-}|}$.

For non-background solutions, the geometries of the temporal surfaces and the transverse spaces do not change. However, the spacetime geometry \eqref{metric} is not of the direct-product form due to the nontrivial metric 1-form ${\ta}$.

\section{Explicit solutions in particular cases}\label{sc:solvable}

\subsection*{Waves and gyratons with vanishing ${\Bt}$}\label{ssc:B0}

As the first explicit example of  higher-dimensional Kundt spacetimes with exact gravitational waves and gyratons we consider a simple case in which the \emph{magnetic field ${\Bt}$ is absent}. Necessarily, it follows from \eqref{MEEB} that the electric field $E$ is uniform, i.e.,
\begin{equation}\label{EBis0}
    E=\text{constant}\;,\quad \Bt=0\;.
\end{equation}
The metric thus takes the form
\begin{equation}\label{metricBis0}
  \tens{g} = \Bigl(\Lambda_-\, r^2 -2 g\, r -2 h\Bigr) \stgrad u\, \stgrad u - \stgrad u \vee \stgrad r  + \stgrad u \vee \ta + \mtr\;,
\end{equation}
where the constant ${\Lambda_-}$, introduced in \eqref{Lambda-}, is now
\begin{equation}\label{Lambda-Bis0}
    \Lambda_- = \frac2{D-2}\, \Lambda  - \frac{D-3}{D-2}\,\kap\epso E^2\;,
\end{equation}
and the Maxwell tensor is
\begin{equation}\label{MaxwellBis0}
  \tens{F} = E\; \stgrad r\wedge \stgrad u+\stgrad u \wedge \tsg \;,
\end{equation}
constrained by the remaining Maxwell equations \eqref{MEsui}, \eqref{MEEBsui}
\begin{eqnarray}
  \grad\tsg \!\!&=&\!\! 0\;,\label{MEsuiBis0}\\
  \div \tsg \!\!&=&\!\!  E \,\div \ta \;.\label{MEEBsuiBis0}
\end{eqnarray}
The Einstein equation \eqref{qequation} for ${\Bt=0}$ reduces to
\begin{equation}\label{EEqBis0}
    \Ric = \Lambda_+ \,\mtr \;,
\end{equation}
where the constant ${\Lambda_+}$ is
\begin{equation}\label{Lambda+Bis0}
    \Lambda_+ = \frac2{D-2}\, \Lambda  + \frac{1}{D-2}\,\kap\epso E^2
\end{equation}
(so that ${\Lambda_+ -\Lambda_- = \kap\epso E^2\ge0}$). The ${(D{-}2)}$-dimensio\-nal Riemannian \emph{transverse space ${N}$ with the metric  ${\mtr}$ can thus be an arbitrary Einstein space}, with its scalar curvature determined by the constant ${\scR=2\Lambda+\kap\epso E^2}$.

The remaining field equations \eqref{aequi}--\eqref{hequi} for ${\ta}$, ${g}$, ${h}$ are
\begin{equation}\label{aequiBis0}
    \frac12 \div \tf  + \Lambda_- \ta -\kap\epso E\, (\tsg -E \ta) - \grad g  =- \tj\;,
\end{equation}
\begin{equation}\label{laplguiBis0}
    \lapl g - \Lambda_- \div \ta = \div \tj\;,
\end{equation}
\begin{equation}\label{hequiBis0}
    \lapl h = \iota - 2 \ta\cdot\grad g +\Lambda_- a^2 - g \div \ta
    + \kap\epso (\tsg {-}E \ta)^2 -\frac12 f^2 \;.
\end{equation}
When we take the exterior derivative $\grad$ of equation \eqref{aequiBis0} we immediately obtain
\begin{equation}\label{grDaequiBis0}
     \frac12 \lapl \tf  + \Lambda_+\, \tf   =- \grad\tj\;,
\end{equation}
while by taking the complementary $\div$ of \eqref{aequiBis0}, equation \eqref{laplguiBis0} for $g$ is recovered.

The problem of finding explicit solutions ${\tens{g}}$, ${\tens{F}}$ which represent gravitational waves and external fields related to a gyratonic source has thus been decoupled, and the Einstein--Maxwell equations can be integrated step by step. First, take any metric ${\mtr}$ which satisfies \eqref{EEqBis0}. Then solve the equation \eqref{grDaequiBis0} for $\tf$ (in the absence of the gyraton it is just the homogeneous Helmholtz equation ${ \lapl \tf  + 2\Lambda_+\, \tf =0}$). This can be subsequently integrated to yield $\ta$ via ${\grad\ta=\tf}$. Clearly, a rotational part\footnote{%
The rotational part of ${\ta}$ is given by the gradient of a potential and therefore its exterior derivative vanishes. See the discussion of potentials following equation \eqref{apotBis0}.}
of ${\ta}$ is not fixed here. However, thanks to the gauge freedom \eqref{rgauge} it can be chosen arbitrarily.

The Maxwell field 1-form  $\tsg$ in \eqref{MaxwellBis0} is then obtained as a solution of equations \eqref{MEsuiBis0} and \eqref{MEEBsuiBis0}. Equation \eqref{laplguiBis0} gives the metric function $g$ and, finally, we find the metric function $h$ by solving the equation \eqref{hequiBis0}.

In fact, this procedure can be made even more explicit if we employ the \emph{potentials corresponding to these quantities}. Namely, the metric 1-form ${\ta}$ can be parameterized using a scalar potential ${\kappa}$ and a co-potential 2-form ${\tl}$, as in \eqref{HodgeTh},
\begin{equation}\label{apotBis0}
    \ta = \grad\kappa - \div\tl\;,
\end{equation}
where we assume the gauge fixing condition ${\grad\tl =0}$ and we ignore a possible harmonic part ${\ta_\harm}$. (Non-trivial harmonics would be important only in compact transverse spaces with special topology. We will include these harmonic terms in the more detailed discussion in the next subsection.) The potentials thus satisfy ${\tf=\grad\ta = -\lapl\tl\,}$ and ${\,\div\ta=\lapl\kappa}$. Analogously, we introduce the potentials for the gyratonic source ${\tj}$:
\begin{equation}\label{jpotBis0}
    \tj = \grad\mu - \div\tn\;.
\end{equation}
Due to the constraint \eqref{MEsuiBis0}, the electromagnetic 1-form ${\tsg}$ can be written in terms of a scalar potential ${\ph}$ only,
\begin{equation}\label{sigpotBis0}
   \tsg = \grad \ph\;.
\end{equation}

Since ${\div\tsg=\lapl\ph}$, the remaining Maxwell equation \eqref{MEEBsuiBis0} takes the form ${\lapl\ph=E \,\div \ta= E\,\lapl\kappa}$, i.e.,
\begin{equation}\label{lappsiEkBis0}
\lapl\bigl(\ph-E\,\kappa\bigr) =0\;.
\end{equation}
This always admits a solution for the electromagnetic potential $\ph$, once the scalar potential $\kappa$ of $\ta$ is known.

Similarly, equation \eqref{laplguiBis0} for the metric function $g$ takes the form
\begin{equation}\label{lapgkmBis0}
\lapl\bigl(g-\Lambda_-\kappa\bigr) = \lapl\mu\;.
\end{equation}
Only the combination ${g-\Lambda_-\kappa}$ is thus determined by the field equations. By inspecting the gauge transformation \eqref{rgauge} we observe that this is the only gauge-invariant combination of ${g}$ and ${\kappa}$. The remaining information in ${g}$ and ${\kappa}$ is gauge-dependent. Therefore ${g}$ or ${\kappa}$ can be chosen arbitrary and, in particular, we can always achieve either ${\kappa=0}$ or ${g=0}$.

Using the relations ${\tf=\grad\ta = -\lapl\tl\,}$ and ${ \grad \tj =  - \lapl\tn }$, we may also rewrite equation \eqref{grDaequiBis0} for $\tf$ as
\begin{equation}\label{grDaequiBis0pot}
\lapl\Bigl(\frac12\lapl\tl+\Lambda_+\,\tl\Bigr)=-\lapl\tn\;.
\end{equation}

The equations for the potentials ${\ph}$, ${\kappa}$, and ${\tl}$ can be integrated, assuming uniqueness of the solution of the Laplace equation. We obtain
\begin{gather}
    \ph = E\,\kappa\;,\\
    g= \Lambda_-\,\kappa +\mu\;\\
    \frac12\lapl\tl+\Lambda_+\,\tl =  - \tn\;,\label{leqB=0}
\end{gather}
where ${\kappa}$ is arbitrary.
Possible pure harmonic contributions to these potentials can be ignored since they are annihilated when evaluating the quantities~${\ta}$ and~${\tsg}$

In particular, pure gravitational waves without a gyratonic source (${\iota=0}$, ${\mu=0}$, ${\tn=0}$), using the gauge ${\kappa=0}$, are given by the metric
\begin{equation}
  \tens{g} =\! \bigl(\Lambda_- r^2 {-}2 h\bigr) \stgrad u\, \stgrad u - \stgrad u \vee \stgrad r  + \stgrad u \vee \ta + \mtr\,.
\end{equation}
The transverse metric ${\mtr}$ solves \eqref{EEqBis0}, ${\ta = - \div \tl}$ with the potential 2-form ${\tl}$ satisfying
\begin{equation}
    \lapl\tl+2\Lambda_+\tl = 0\;,
\end{equation}
and the metric function ${h}$ satisfies
\begin{equation}
    \lapl h = \Lambda_+\, a^2 - 2 \Lambda_+^2\, \lambda^2\;.
\end{equation}
Here, the constants ${\Lambda_-}$, ${\Lambda_+}$ are given by \eqref{Lambda-Bis0}, and \eqref{Lambda+Bis0}, respectively. 
The electric field is just ${\tens{F} = E\; \stgrad r\wedge \stgrad u}$.

\subsection*{Waves and gyratons on any direct-product transverse space}\label{ssc:dpa}

In this section we switch on the background magnetic field. However, since the 2-form ${\Bt}$ must be harmonic, we expect that it can be nontrivial only in spaces with a special topology.\footnote{%
For a compact transverse space, a space of nontrivial harmonic 2-forms is equivalent to the second cohomology group. In a noncompact case we have a analogous relation if we assume reasonably restrictive boundary conditions for the background field.}

\subsubsection*{The direct product ansatz}

An important example of such topologically special spaces is a transverse ${(D{-}2)}$-dimensional space ${N}$ which has a direct product structure
\begin{equation}\label{dpN}
    N = \dpc{0}N\times\dpc{1}N\times \dpc{2}N\times\cdots\;,
\end{equation}
where the components ${\dpc{K}N}$, ${K=1,2,\dots}$ are 2-dimensional while ${\dpc{0}N}$ is an (optional) exceptional component of an arbitrary dimension on which the magnetic field vanishes. The direct-product structure is reflected also in the metric which takes a simple orthogonal form
\begin{equation}\label{dpmetric}
    \mtr = \sum_{K=0,1,\dots} \!\! \dpc{K}\mtr\;.
\end{equation}
The metrics ${\dpc{K}\mtr}$ act only on the vectors tangent to the ${K}$-th component and they are independent of a position in any other ${L}$-th component ${\dpc{L}N}$, ${L\neq K}$.

The harmonic 2-form ${\Bt}$ can be found using the ansatz
\begin{equation}\label{dpB}
   \Bt=\sum_{K=1,2,\dots} \!\!\dpc{K}B\; \dpc{K}\lc\;,
\end{equation}
where ${\dpc{K}\lc}$ is the Levi-Civita 2-form of the metric ${\dpc{K}\mtr}$ in the 2-dimensional component ${\dpc{K}N}$, and each ${\dpc{K}B}$ is a scalar.
The harmonic conditions \eqref{EBharm} now enforce the uniformity of the background electromagnetic field, namely that the component ${E}$ and the scalars ${\dpc{K}B}$ must be transverse-space constants:
\begin{equation}\label{EBconst}
    E=\text{constant}\;,\quad \dpc{K}B = \text{constant}\;.
\end{equation}
Consequently, the magnetic 2-form ${\Bt}$ is covariantly constant on the transverse space~${N}$,
\begin{equation}\label{covdB0}
    \covd\Bt=0\;.
\end{equation}
We also obtain
\begin{equation}\label{Bquad}
    \BBt = \sum_{K=1,2,\dots} \dpc{K}B^2\; \dpc{K}\mtr\;,\quad B^2 = \sum_{K=1,2,\dots} \dpc{K}B^2\;.
\end{equation}

Clearly, the case discussed in the previous section corresponds to the trivial choices
\begin{equation}
    \dpc{K}B=0\;,
\end{equation}
with possibly only the ${\dpc{0}N}$ component present and the \mbox{2-dimensional} components missing.

We also assume an adjusted structure for the 1-forms ${\ta}$, ${\tsg}$, and ${\tj}$, as given below.

\subsubsection*{Equation for the transverse metric ${\mtr}$}

A straightforward consequence of the direct-product ansatz \eqref{dpmetric} is that  both the Riemann and Ricci tensors have the similar structure
\begin{equation}\label{dpRRic}
    \Riem =  \sum_{K=0,1,\dots} \!\!\!\!\dpc{K}\Riem\;,\quad
    \Ric =  \sum_{K=0,1,\dots} \!\!\!\!\dpc{K}\mspace{2mu}\Ric\;,
\end{equation}
with ${\dpc{K}\Riem}$ and  ${\dpc{K}\mspace{2mu}\Ric}$ depending only on a position in the \mbox{${K}$-th} component ${\dpc{K}N}$, see e.g.~\cite{Ficken:1939}. The trace-free transverse stress-energy tensor \eqref{tracefreeT} reads
\begin{equation}
   \frac1{\epso}\,\tT_{\trpr\trpr}^{\,\TF}=\sum_{K=1,2,\dots} \Bigl(\dpc{K}B^2-\frac{2}{D{-}2}B^2\Bigr)\dpc{K}\mtr\;.
\end{equation}
Substituting this into \eqref{EEq} we obtain the equation which can be split into the orthogonal components tangent to ${\dpc{K}N}$, namely,
\begin{equation}\label{compEEq}
    \dpc{K}\mspace{2mu}\Ric = \dpc{K}\Lambda_+\,\dpc{K}\mtr\;,
\end{equation}
${K=0,1,2,\dots}$, where the constants ${\dpc{K}\Lambda_+}$  are now given~as
\begin{equation}\label{lambda+}
    \dpc{K}\Lambda_+ = \frac{2}{D{-}2}\bigl(\Lambda +\rho\bigr)+\kap\epso\Bigl(\dpc{K}B^2-\frac{2}{D{-}2}B^2\Bigr)\;,
\end{equation}
and we set ${\dpc{0}B=0}$.

The equations \eqref{compEEq} imply that \emph{each component ${\dpc{K}N}$ is an Einstein space}. For the 2-dimensional components ${K=1,2,\dots}$ it means that the metric ${\dpc{K}\mtr}$ is the standard homogeneous metric of a sphere ${S_2}$, plane ${E_2}$, or hyperbolic plane ${H_2}$, depending on the sign of its Gaussian curvature, which is given by ${\dpc{K}\Lambda_+}$.

The zeroth component ${\dpc{0}N}$ (whose ${\dpc{0}B=0}$) must also be an Einstein space, but for higher dimensions its geometry is not fixed uniquely. For the maximally symmetric choice of ${\dpc{0}N}$ see the work \cite{CardosoDiasLemos:2004}. On the other hand, if the zeroth component ${\dpc{0}N}$ is just \mbox{1-dimensional}, its curvature vanishes, ${\dpc{0}\Lambda_+=0}$. This implies the restriction ${\Lambda+\tau=0}$ and consequently ${\dpc{K}\Lambda_+=\kap\epso\,\dpc{K}B^2}$.

Combining any transverse-space product of ${\dpc{0}N}$, ${S_2}$, ${E_2}$, or ${H_2}$ of curvatures ${\dpc{K}\Lambda_+}$ with the temporal surface ${T}$ of constant curvature ${\Lambda_-}$, see \eqref{GaussTemp}, we obtain a great number of higher-dimensional generalizations of (anti-)Nariai, Bertotti--Robinson, Pleba\'{n}ski--Hacyan and Minkowski spaces which are well-known direct-product 4-dimensional spacetimes \cite{Stephani:2003,GriffithsPodolsky:2009,KadlecovaZelnikovKrtousPodolsky:2009}.

\subsubsection*{Potentials}

As in the previous case ${\Bt=0}$ it is very useful to introduce the potentials for the 1-forms ${\ta}$ and ${\tj}$. However, for completeness, we will now also include harmonic terms in their Hodge decompositions \eqref{HodgeTh}.

The metric 1-form ${\ta}$ may be parameterized using a scalar potential ${\kappa}$ and a co-potential 2-form ${\tl}$ as in \eqref{HodgeTh},
\begin{equation}\label{apot}
    \ta = \grad\kappa - \div\tl + \ta_\harm\;,
\end{equation}
where ${\grad\tl =0}$ and ${\ta_\harm}$ is a harmonic 1-form. These potentials satisfy
\begin{equation}\label{apotprop}
    \grad\ta = -\lapl\tl\;,\quad \div\ta=\lapl\kappa\;.
\end{equation}

Now we can explicitly formulate the condition that the metric 1-form ${\ta}$ is adjusted to the direct-product structure. We will assume that the co-potential ${\tl}$ has a similar structure as ${\Bt}$ given in \eqref{dpB},
\begin{equation}\label{dpl}
    \tl = \dpc{0}\tl\; +\!\! \sum_{K=1,2,\dots} \!\!\!\dpc{K}\lambda\; \dpc{K}\lc\;,
\end{equation}
The co-potential gauge condition ${\grad\tl=0}$ then implies ${\grad\dpc{K}\lambda\wedge\dpc{K}\lc=0}$ and ${\grad\,\dpc{0}\tl=0}$, which means that the scalars ${\dpc{K}\lambda}$ only depend on a position in the component ${\dpc{K}N}$, and ${\dpc{0}\tl}$ on a position in ${\dpc{0}N}$.
The 2-form ${\tf=\grad\ta}$ inherits a similar structure
\begin{gather}
    \tf = \dpc{0}\tf  \;+\!\! \sum_{K=1,2,\dots} \!\!\!\dpc{K}f\,\dpc{K}\lc\;,\label{fdp}\\
    \dpc{0}\tf=-\lapl\dpc{0}\tl\;,\quad
    \dpc{K}f=-\lapl\dpc{K}\lambda\;.\label{fdpcomp}
\end{gather}

The divergent part ${-\div\tl}$ of the metric 1-form ${\ta}$ can be written as
\begin{equation}\label{divldp}
    -\div\tl = -\div\dpc{0}\tl \;+\!\! \sum_{K=1,2,\dots} \!\!\!\dpc{K}\lc\cdot\grad\dpc{K}\lambda\;,
\end{equation}
which means that the projection on the directions tangent to the ${K}$-th component depends only on a position in this component.

We require a similar property to be valid also for the harmonic part~${\ta_\harm}$,
\begin{equation}\label{aharmdp}
    \ta_\harm = \sum_{K=0,1,\dots} \!\!\!\dpc{K}\ta_\harm\;,
\end{equation}
with {${\dpc{K}\ta_\harm}$ depending only on a position in ${\dpc{K}N}$} and harmonic on this component,\footnote{%
Thanks to the orthogonal character of the metric and our assumption about the spatial dependence of ${\ta_\harm}$, we do not have to distinguish here the exterior derivative ${\grad}$ and the divergence ${\div}$ on the whole space ${N}$ and its components ${\dpc{K}N}$.}
\begin{equation}\label{aharmdpcond}
    \grad \dpc{K}\ta_\harm = 0\;,\quad \div\dpc{K}\ta_\harm=0\;.
\end{equation}
Note, however, that the spatial dependence of the rotational part ${\grad\kappa}$ of the metric 1-form ${\ta}$ is not restricted in a similar way as the divergent part ${-\div\tl}$ and the harmonic ${\ta_\harm}$. We will see shortly that this part can be solved explicitly in general, without any assumptions on its direct-product structure.

Let us also mention a useful property of the co-potential ${\tl}$. Using the fact that the \mbox{2-dimensional} metrics ${\dpc{K}\mtr}$ are maximally symmetric, with the curvature tensors ${\dpc{K}\mspace{3mu}\Ric}$ and ${\dpc{K}\Riem}$ expressed just in terms of ${\dpc{K}\mtr}$, and using its direct-product form \eqref{dpl}, the Weitzenb\"ock--Bochner identity \eqref{WBident} reduces to
\begin{equation}\label{lapllambda}
    \lapl\tl=\LB\tl\;.
\end{equation}

Finally, we can introduce the potentials for the gyratonic source ${\tj}$. We assume a similar structure as for~${\ta}$,
\begin{equation}\label{jpot}
    \tj = \grad\mu - \div\tn + \tj_\harm\;,
\end{equation}
with ${\grad\tn=0}$ and
\begin{equation}\label{dpn}
    \tn = \dpc{0}\tn\; +\!\! \sum_{K=1,2,\dots} \dpc{K}\nu\; \dpc{K}\lc\;,
\end{equation}
in which ${\dpc{K}\nu}$ depend only on a position in ${\dpc{K}N}$, and ${\dpc{0}\tl}$ on a position in ${\dpc{0}N}$. The harmonic ${\tj_\harm}$ splits into the harmonics on each component,
\begin{equation}\label{jharmdp}
    \tj_\harm \,=\!\! \sum_{K=0,1,\dots}\!\!\!\! \dpc{K}\tj_\harm\;,\quad \grad\dpc{K}\tj=0\;,\quad\div\dpc{K}\tj=0\;.
\end{equation}

Now we can explicitly integrate the field equations.

\subsubsection*{Electromagnetic 1-form ${\tsg}$}

The electromagnetic 1-form ${\tsg}$ satisfies condition \eqref{MEsui} which implies the existence of a scalar potential ${\ph}$,
\begin{equation}\label{sigpot}
   \tsg = \grad \ph + \tsg_\harm\;,
\end{equation}
with ${\tsg_\harm}$ being a harmonic 1-form. Clearly ${\div\tsg=\lapl\ph}$.

Taking into account the definition \eqref{apot}, the identities \eqref{apotprop}, \eqref{lapllambda}, \eqref{covdB0}, and relation\footnote{%
This relation holds for any harmonic ${\Bt}$ and 1-form ${\tens{\alpha}}$. Indeed, let ${\tens{\beta}}$ be a ${(D-4)}$-form dual to ${\Bt}$, that is ${\Bt=*\tens{\beta}}$. Then
${\div\bigl(\tens{\alpha}\cdot\Bt\bigr)
  =\div*\bigl(\tens{\beta}\wedge\tens{\alpha}\bigr)
  =(-1)^{D{-}3}*^{\!-1}\grad\bigl(\tens{\beta}\wedge\tens{\alpha}\bigr)}$. Using the Leibnitz rule and ${\grad\tens{\beta}=0}$ we obtain
$\div\bigl(\tens{\alpha}\cdot\Bt\bigr)=-*\bigl(\tens{\beta}\wedge\grad\tens{\alpha}\bigr)=-\Bt\bullet\grad\tens{\alpha}$.
}
${\div\bigl(\tens{\alpha}\cdot\Bt\bigr) = -\Bt\bullet\grad\tens{\alpha}}$, the Maxwell equation \eqref{MEEBsui} leads to the explicit equation for the potential ${\ph}$,
\begin{equation}\label{pheq}
    \lapl\bigl(\ph-E\,\kappa-\Bt\bullet\tl\bigr)=0\;.
\end{equation}
Ignoring the possibility of nontrivial scalar harmonics in the noncompact case (and neglecting trivial additive constants), it can be solved as
\begin{equation}\label{phsol}
    \ph=E\,\kappa+\Bt\bullet\tl = E\,\kappa+\sum_{K=1,2,\dots}\!\!\!\dpc{K}B\,\dpc{K}\lambda\;.
\end{equation}

For the 1-form ${\tsg}$ we thus obtain
\begin{equation}\label{sgsol}
    \tsg=E\,\grad\kappa+\sum_{K=1,2,\dots}\!\!\!\dpc{K}B\,\grad\dpc{K}\lambda+\tsg_\harm\;.
\end{equation}
We assume that the harmonic part ${\tsg_\harm}$ splits into the corresponding harmonic 1-forms on each component ${\dpc{K}N}$, similarly as for ${\ta_\harm}$,
\begin{equation}\label{sgharmdp}
    \tsg_\harm = \sum_{K=0,1,\dots} \!\!\!\dpc{K}\tsg_\harm\;,
\end{equation}
with {${\dpc{K}\tsg_\harm}$ depending only on a position in ${\dpc{K}N}$} and satisfying
\begin{equation}\label{sgharmdpcond}
    \grad \dpc{K}\tsg_\harm = 0\;,\quad \div\dpc{K}\tsg_\harm=0\;.
\end{equation}

\subsubsection*{Rotational part of ${\ta}$ and the metric function ${g}$}

Now we analyze the equations for the metric \mbox{1-form} ${\ta}$. First we substitute the potentials \eqref{apot} and \eqref{jpot} into \eqref{laplgui}, obtaining
\begin{equation}\label{lapgkm}
\lapl\bigl(g-\Lambda_-\kappa\bigr) = \lapl\mu\;.
\end{equation}
Ignoring trivial additive constants (and possible scalar harmonics in the noncompact case) we solve this equation explicitly as
\begin{equation}\label{gkm}
    g-\Lambda_-\kappa = \mu\;.
\end{equation}

We observe that the metric function ${g}$ is closely related to the potential ${\kappa}$. Only their difference is determined by the potential ${\mu}$ of the gyraton source ${\tj}$. This has a clear physical reason: both ${g}$ and ${\kappa}$ change under gauge transformation \eqref{rgauge}, namely
\begin{equation}\label{gkappargauge}
    \tilde g = g + \Lambda_- \psi\;,\quad\tilde \kappa =\kappa+\psi\;.
\end{equation}
Therefore, they cannot be uniquely determined by the field equations. However, their combination ${g-\Lambda_-\kappa}$ is gauge invariant and uniquely given by the source ${\mu}$. We can actually achieve any particular form of ${g}$ (or of ${\kappa}$, respectively) by choosing  a suitable gauge transformation. Useful gauge fixing conditions are ${g=0}$, or ${\kappa=0}$, or ${\ph=0}$ (if ${E\neq0}$, cf.\ eq.~\eqref{phsol}).

\subsubsection*{Divergent part of ${\ta}$}

The potential ${\kappa}$ determines the rotational part of ${\ta}$, cf.\ eq.~\eqref{apot}. The equation for the divergent part of ${\ta}$ (encoded in the co-potential ${\tl}$, or also in the 2-form ${\tf}$, cf.~\eqref{fdef}) can be extracted by taking exterior derivative of the equation \eqref{aequi}:
\begin{equation}\label{daeq1}
    \frac12\lapl\tf+(\Lambda_-{+}\kap\epso E^2)\tf-\kap\epso\grad\bigl[(\tsg{-}E\ta)\cdot\Bt\bigr]=-\grad\tj\;.
\end{equation}
Substituting for ${\ta}$, ${\ts}$, and ${\tj}$ the corresponding expressions with potentials, a rather involved calculation then leads to
\begin{equation}\label{daeq2}
    \frac12\lapl\tf+\bigl(\Lambda_-\mtr+\kap\epso (E^2\mtr{+}\BBt)\bigr)\cdot\tf = \lapl\tn\;.
\end{equation}
Here we have repeatedly used the block structures \eqref{dpB}, \eqref{dpl} of ${\Bt}$ and ${\tl}$, which imply
\begin{equation*}
  (\div\tl)\cdot\Bt=-\sum_K\dpc{K}B\,\grad\dpc{K}\lambda \;,
\end{equation*}
i.e., ${\grad\bigl((\div\tl)\cdot\Bt\bigr)=0}$, and
\begin{equation*}
\Bt\cdot\sum_K\dpc{K}B\,\grad\dpc{K}\lambda=-\sum_K \dpc{K}B^2\div(\dpc{K}\lambda\dpc{K}\lc)\;,
\end{equation*}
so that
${\grad\bigl(\Bt\cdot\sum_K\dpc{K}B\,\grad\dpc{K}\lambda) = -\BBt\cdot\lapl\tl = \BBt\cdot\tf}$. The assumptions \eqref{aharmdpcond}, \eqref{sgharmdpcond} on the harmonic parts of ${\ta_\harm}$ and ${\tsg_\harm}$ also guarante that these harmonic terms do not contribute to \eqref{daeq2}.

Splitting \eqref{daeq2} into independent parts tangent to each component we obtain the equations for ${\dpc{0}\tf}$ and ${\dpc{K}f}$:
\begin{equation}\label{feqdp0}
    \frac12\lapl\dpc{0}\tf+\dpc{0}\Lambda_+\dpc{0}\tf = \lapl\dpc{0}\tn \;,
\end{equation}
and
\begin{equation}\label{feqdp}
    \frac12\lapl\dpc{K}f+\dpc{K}\Lambda_+\dpc{K}f = \lapl\dpc{K}\nu\;,
\end{equation}
${K=1,2,\dots}$, with
\begin{equation}\label{lambda+-rel}
    \dpc{K}\Lambda_+ = \Lambda_- + \kap\epso(E^2+\dpc{K}B^2)\;,
\end{equation}
which is equivalent to the relation \eqref{lambda+}.

Equation~\eqref{feqdp0} repeats the equation \eqref{grDaequiBis0pot} for the case ${\Bt=0}$ (recall that ${\tf=-\lapl\tl}$). The equations \eqref{feqdp} are analogous equations for the scalar quantities ${\dpc{K}f}$ on each 2-dimensional component ${\dpc{K}N}$. All these equations have the linear form of the Helmholz equation, which can be solved in terms of the Green functions. For maximally symmetric 2-dimensional components ${\dpc{K}N}$ these can be written down explicitely (see, e.g., \cite{KadlecovaZelnikovKrtousPodolsky:2009}). It is important to observe that ${\dpc{K}\Lambda_+}$ is the Gaussian curvature of the \mbox{2-metric} on ${\dpc{K}N}$, ${K=1,2,\dots}$, which determines the scalar Laplace operator in equation \eqref{feqdp}. For positive values ${\dpc{K}\Lambda_+>0}$, the geometry is spherical and the coefficient in front of the linear term is related to the eigenvalue of this Laplace operator. The equation has thus a solution even in the absence of sources.

After solving the Helmholz equations for the components of ${\tf}$, the potentials ${\dpc{K}\lambda}$ can then be obtained by solving the Poisson equations \eqref{fdpcomp}. Alternatively, we can also substitute \eqref{fdpcomp} into \eqref{feqdp0} and \eqref{feqdp}. Integration of the outer Laplace operator gives the Helmholz equations
\begin{gather}
    \frac12\lapl\dpc{0}\tl+\dpc{0}\Lambda_+\dpc{0}\tl = -\dpc{0}\tn \;,\label{leqdp0}\\
    \frac12\lapl\dpc{K}\lambda+\dpc{K}\Lambda_+\dpc{K}\lambda = -\dpc{K}\nu\;,\label{leqdp}
\end{gather}
cf.\ equation~\eqref{leqB=0}.

\subsubsection*{Harmonic parts  ${\ta_\harm}$ and ${\tsg_\harm}$}

The harmonic forms ${\ta_\harm}$ and ${\tsg_\harm}$ do not enter the equations \eqref{lapgkm} and \eqref{feqdp}, but they cannot be ignored in the original field equation \eqref{aequi}. Substituting here the expressions for potentials \eqref{apot}, \eqref{jpot} and \eqref{sgsol}, and using the field equations \eqref{gkm} and \eqref{leqdp}, we obtain the constraint for these harmonics as
\begin{equation}\label{harmeq}
    \Lambda_-\ta_\harm - \kap\epso(\tsg_\harm-E\ta_\harm)\cdot(E\mtr+\Bt) + \tj_\harm = 0\;.
\end{equation}
Contracting with ${(E\mtr-\Bt)}$ from the right, using the antisymmetry of ${\Bt}$ and definition \eqref{BBdef}, we obtain
\begin{equation}\label{tsghrm}
    \kap\epso(E^2\mtr+\BBt)\cdot(\tsg_\harm - E\,\ta_\harm) = (E\mtr+\Bt)\cdot(\Lambda_-\ta_\harm+\tj_\harm)\;.
\end{equation}
Splitting onto the components ${\dpc{K}N}$ we thus obtain
\begin{equation}\label{Ktsghrm}
    \dpc{K}\tsg_\harm = E\,\dpc{K}\ta_\harm +\frac{E\,\dpc{K}\mtr+\dpc{K}B\,\dpc{K}\lc}{\kap\epso(E^2+\dpc{K}B^2)}\cdot(\Lambda_-\dpc{K}\ta_\harm+\dpc{K}\tj_\harm)\;.
\end{equation}
This expresses the electromagnetic harmonic 1-form~${\tsg_\harm}$ in terms the harmonic parts of the metric 1-form ${\ta}$ and the source ${\tj}$.

\subsubsection*{Equation for  ${h}$}

The only remaining quantity is the metric scalar function ${h}$. It is determined by the equation \eqref{hequi} where all the terms on the right-hand side can be obtained after solving all the field equations discussed above. We thus obtain the Poisson equation, albeit with a rather complicated source.

\subsection*{Character of the background geometries for the waves and gyratons}

\subsubsection*{Direct-product spacetimes}

Let us shortly summarize and discuss the background geometries of the spacetimes studied. Clearly, the geometries given by \eqref{metricbkgr}--\eqref{Maxwellbkgr} with the spatial part ${\mtr}$ satisfying \eqref{EEqBis0}, or \eqref{dpmetric} with \eqref{compEEq}, are generalizations of \mbox{4-dimensional} direct-product spacetimes, see, e.g., \cite{Stephani:2003,GriffithsPodolsky:2009}.

In ${D=4}$, the metric of such spacetimes is given by the sum of a homogeneous Lorentzian \mbox{2-metric} on ${T}$ with a homogeneous Riemannian \mbox{2-metric} on ${N}$. Depending on the signs of the curvatures of these components this yields vacuum (anti-)Nariai and Minkowski, or electrovacuum Bertotti--Robinson and Pleba\'{n}ski--Hacyan spacetimes. The waves and gyratons on these backgrounds were discussed in detail in our previous work \cite{KadlecovaZelnikovKrtousPodolsky:2009}.

We observe that the higher-dimensional case studied here generalizes these spacetimes in two natural ways: (i) It represents a direct-product spacetime with one 2-dimensional homogeneous Lorentzian  component and one Riemannian component of a general dimension ${D{-}2}$ with the Einstein geometry and vanishing magnetic field. (ii) It is a direct-product spacetime composed from one Lorentzian and several Riemannian 2-dimensional homogeneous components with independent uniform magnetic fields associated with each of the spatial component. Of course, we also admit a straightforward combination of both these cases. Notice that the case (i) generalizes the particular forms of (anti-)Narai and Bertotti--Robinson spacetimes discussed in \cite{CardosoDiasLemos:2004} where ${N=E_{D{-}2}}$, ${S_{D{-}2}}$, or ${H_{D{-}2}}$.

Inspecting relation \eqref{lambda+-rel} we find that the Gaussian curvatures ${\dpc{K}\Lambda_+}$ of the spatial 2-dimensional components ${\dpc{K}N}$ must be greater or equal then the Gaussian curvature ${\Lambda_-}$ of the temporal surface ${T}$. They are equal to ${\Lambda_-}$ if, and only if, the electric field ${E}$ and the corresponding magnetic component ${\dpc{K}B}$ vanish simultaneously. The richest family of spacetimes is thus obtained if the temporal surface has the anti-de~Sitter geometry, ${\Lambda_-<0}$. In such a case the spatial components can be either spheres ${S_2}$, planes ${E_2}$, or hyperbolic planes ${H_2}$. For the flat temporal surface, ${\Lambda_-=0}$, all spatial components must be spheres ${S_2}$, except the case in which the electric and magnetic fields are missing, implying flat spatial components. Indeed, for ${\Lambda_-=0}$, all the spatial Gaussian curvatures read ${\dpc{K}\Lambda_+=\kap\epso(E^2+\dpc{K}B^2)}$. Finally, for the de~Sitter temporal surface, ${\Lambda_->0}$, the spatial components must be spheres.

Recall that in any dimension ${D\ge4}$ these direct-product spacetimes are of algebraic type D or are conformally flat, cf.~\eqref{Weylframecomp}.

\subsubsection*{K\"ahler structure}

An important subcase is obtained when the dimension ${D{-}2}$ is even, the exceptional zeroth component is missing, and all ${\frac{D-2}{2}}$ components of the background magnetic field are the same,
\begin{equation}\label{Bcompsame}
  \dpc{K}B=b\;,
\end{equation}
with the constant ${b}$ given by ${b^2=\frac2{D{-}2}\,B^2}$.
In such a case ${\BBt=b^2\,\mtr}$. The magnetic field ${\Bt}$ \eqref{dpB} can thus be renormalized to become a K\"ahler 2-form ${\tens{\omega}}$,
\begin{equation}\label{Kahler2form}
    \tens{\omega}=\frac1b\,\Bt = \sum_{K=1,2\dots} \dpc{K}\lc\;,
\end{equation}
compatible with the transverse metric ${\mtr}$ given by \eqref{dpmetric}. Indeed, thanks to equation \eqref{covdB0} and the fact that ${b=\text{constant}}$, the 2-form ${\tens{\omega}}$ equips the transverse space ${N}$ not only with an almost K\"ahler structure but also with the K\"ahler structure.

Notice, that the assumption \eqref{Bcompsame} implies that the trace-free stress-energy tensor \eqref{tracefreeT} vanishes,
\begin{equation}\label{almostKahler}
   \tT^\TF_{\trpr\trpr}=0\;.
\end{equation}
Such a condition thus guarantees a considerable simplification of the transverse part of the Einstein equations \eqref{EEq}, even without assuming the direct-product structure of the transverse manifold made in \eqref{dpN}. Moreover, the Maxwell equations \eqref{trMaxwell} imply that the 2-form \eqref{Kahler2form} provides the almost K\"ahler structure for  the transverse space ${N}$. Interestingly, such solutions are analogous to expanding Robinson--Trautman electrovacuum spacetimes in ${D\ge4}$ \cite{OrtaggioEtal:2008}.

It posses an interesting open question whether there exist backgrounds outside the class discussed in Sec.~\ref{ssc:dpa}, i.e., those without the direct-product structure, which would still satisfy the assumption \eqref{almostKahler} and for which it would be possible to  separate the remaining field equations. The main difficulty here lies in the separation of the equation for the divergent part of the metric \mbox{1-form}~${\ta}$.

\section{Summary}\label{sc:sum}

We presented a new large class of gyraton solutions of the Einstein--Maxwell
equations in higher dimensions, in particular the gyratons on
type D or conformally flat background spacetimes which are formed as a direct product of
constant-curvature 2-spaces (and a possible
additional Einstein space of an arbitrary dimension in which the magnetic
field vanishes).
In four dimensions these background spacetimes involve the famous (anti-)Nariai,
Bertotti--Robinson, and Pleba\'nski--Hacyan spacetimes. The
background geometries are solutions of the Einstein--Maxwell
equations corresponding to the uniform background
electric and magnetic fields. These gyraton solutions belong to the Kundt family
of shearfree, twistfree and nonexpanding spacetimes.

Gyratons describe the gravitational field created by a
stress-energy tensor of a spinning (circularly polarized)
high-frequency beam of radiation of electromagnetic, neutrino,
or any other massless fields. They also provide
a good approximation for the gravitational field of a
beam of ultrarelativistic particles with an intrinsic spin. The gyratons
generalize standard vacuum or pure radiation pp-waves or Kundt waves by
admitting a nonzero angular momentum of the source.
This leads to nontrivial ${\scriptstyle u\trpr}$-components of the Einstein
equations, in addition to the pure radiation ${\scriptstyle uu}$-component which appears for
the simplest pp-waves and Kundt waves.
We have shown that all the Einstein--Maxwell equations
can be solved explicitly for any distribution of such
matter sources. Namely, the task has been reduced to a linear problem involving scalar Green
functions on separate transverse components,
which are typically 2-dimensional sphere, plane or hyperboloid.

\begin{acknowledgments}
P.~K. was supported by Grant GA\v{C}R~202/09/0772, and J.~P. by Grant GA\v{C}R~P203/12/0118.
A.~Z. was supported by the Natural Sciences and Engineering Research Council of Canada and by
the Killam Trust. A.~Z. is also grateful to the Charles University in Prague for hospitality during his work on this paper.
H.~K. was supported by Grant GA\v{C}R~205/09/H033.
The authors are grateful to Valeri Frolov for stimulating discussions.
\end{acknowledgments}

\appendix

\section{Gauge freedom}\label{apx:gauge}

The construction of the transverse spaces and splitting of the metric in the form \eqref{metric} is not unique. It contains three partially ambiguous choices---gauges. The choice of the coordinate ${u}$ ($u$-gauge), the choice of the coordinate ${r}$ ($r$-gauge), and the choice of the flow ${\tens{w}}$ ($w$-gauge).

The foliation of null hypersurfaces ${S}$ defines the coordinate ${u}$ up to a reparametrization
\begin{equation}\label{uresc}
  u \rightarrow \tilde{u}=f(u)\;,
\end{equation}
with one-to-one function ${f}$ of one variable. The reparametrization has to be accompanied by rescaling of ${\tens{k}}$ and ${r}$, however, the transverse foliation ${N}$ and the temporal surfaces ${T}$ are unchanged. Different quantities transform as follows:
\begin{equation}\label{ugauge}
\begin{gathered}
  \tilde{u} = f\;,\quad
  \tilde{r} = \frac1{f'}r\;,\\
  \tilde{\tens{k}} = f' \tens{k}\;,\quad
  \tilde{\tens{w}} = \frac1{f'}\Bigl(\tens{w}+\frac{f''}{f'}r\tens{k}\Bigr)\;,\\
  \tilde{H} = \frac1{f'{}^2}\Bigl(H+\frac{f''}{f'}r\Bigr)\;,\quad
  \tilde{\ta} = \frac1{f'} \ta\;,\quad
  \tilde{\mtr} = \mtr\;.
\end{gathered}
\end{equation}
Clearly, the ${u}$-gauge changes both derivatives along ${\tens{k}}$ and~${\tens{w}}$.

The ${r}$-gauge freedom is related to the fact that the affine parameter of the geodesic generated by ${\tens{k}}$ is defined up to a constant. The coordinate function ${r}$ is thus defined up to the ${r}$-independent shift:
\begin{equation}\label{rshift}
  r \rightarrow \tilde{r}=r+\psi\;,\quad \rder \psi = 0\;.
\end{equation}
Such a shift changes the transverse space foliation, although the typical transverse space ${N}$ is unchanged. The \mbox{${r}$-gauge} thus modifies the embedding ${\iota_{u,r}}$ of ${N}$ into spacetime ${M}$. If we additionally require that the temporal foliation ${T}$ remains unchanged, we find that the change of ${r}$ has to be accompanied by:
\begin{equation}\label{rgauge}
\begin{gathered}
  \tilde{u} = u\;,\quad
  \tilde{r} = r+\psi\;,\\
  \tilde{\tens{k}} = \tens{k}\;,\quad
  \tilde{\tens{w}} = \tens{w} - \uder\psi \tens{k}\;,\\
  \tilde{H} = H-\uder\psi\;,\quad
  \tilde{\ta} = \ta+\grad\psi\;,\quad
  \tilde{\mtr} = \mtr\;,
\end{gathered}
\end{equation}
where ${\grad\psi}$ represents the gradient on the transverse space, i.e., only transverse components of the spacetime gradient ${\stgrad\psi}$.
We also see that the ${r}$-gauge changes just the derivative along ${\tens{w}}$.

Finally, the ${w}$-gauge leaves the transverse spaces ${N}$ unchanged but changes the identification of them for different values of the coordinate ${u}$. When restricted to the typical transverse space ${N}$, it can be viewed as the ${u}$-dependent (and ${r}$-independent) family of diffeomorphisms of ${N}$. The generator of this family of diffeomorphisms is exactly the vector field ${\tens{\xi}}$ by which the ${w}$-gauge modifies the flow ${\tens{w}}$ \cite{KrtousPodolsky:inprep}:
\begin{equation}\label{wshift}
  \tens{w} \rightarrow \tilde{\tens{w}}=\tens{w}+\tens{\xi}\;,\quad \rder {\tens{\xi}} =0\;.
\end{equation}
The corresponding changes of the metric quantities are\\
\begin{equation}\label{wgauge}
\begin{gathered}
  \tilde{u} = u\;,\quad
  \tilde{r} = r\;,\\
  \tilde{\tens{k}} = \tens{k}\;,\quad
  \tilde{\tens{w}} = \tens{w} + \tens{\xi}\;,\\
  \tilde{H} = H-\ta\cdot\tens{\xi}-\frac12\xi^2\;,\quad
  \tilde{\ta} = \tens{a}+\tens{\xi}\;,\quad
  \tilde{\mtr} = \mtr\;.
\end{gathered}
\end{equation}
As for the ${r}$-gauge, only the ${u}$-derivative is modified.

It is explained in more detail in \cite{KrtousPodolsky:inprep} that although both the transverse spaces ${N}$ and the temporal planes ${T}$ are not determined uniquely by the spacetime geometry, the assumption ${\rder{\ta}=0}$ of the existence of the temporal planes orthogonal to the transverse spaces is independent of a specific choice of the gauge, in particular of the choices of ${N}$.


\begin{thebibliography}{60}
\expandafter\ifx\csname natexlab\endcsname\relax\def\natexlab#1{#1}\fi
\expandafter\ifx\csname bibnamefont\endcsname\relax
  \def\bibnamefont#1{#1}\fi
\expandafter\ifx\csname bibfnamefont\endcsname\relax
  \def\bibfnamefont#1{#1}\fi
\expandafter\ifx\csname citenamefont\endcsname\relax
  \def\citenamefont#1{#1}\fi
\expandafter\ifx\csname url\endcsname\relax
  \def\url#1{\texttt{#1}}\fi
\expandafter\ifx\csname urlprefix\endcsname\relax\def\urlprefix{URL }\fi
\providecommand{\bibinfo}[2]{#2}
\providecommand{\eprint}[2][]{\url{#2}}


\bibitem{Tolman:1934}R. C. Tolman, {\it Relativity, Thermodynamics and Cosmology} (Clarendon Press, Oxford, 1934).
\bibitem{Peres:1960} A. Peres, Phys. Rev. {\bf 118}, 1105 (1960).
\bibitem{Bonnor:1969a} W. B. Bonnor, Commun. Math. Phys. {\bf 13}, 163 (1969).
\bibitem{Bonnor:1969b} W. B. Bonnor, Int. J. Theor. Phys. {\bf 2}, 373 (1969).
\bibitem{Bonnor:1970a} W. B. Bonnor, Int. J. Theor. Phys. {\bf 3}, 57 (1970).
\bibitem{Brinkmann:1925}H. W. Brinkmann, Math. Ann. {\bf 94}, 119 (1925).
\bibitem{Stephani:2003} H. Stephani, D. Kramer, M. MacCallum, C. Hoenselaers, and E. Herlt, {\it Exact Solutions of Einstein's Field Equations} (Cambridge University Press, Cambridge, 2003).
\bibitem{GriffithsPodolsky:2009} J. B. Griffiths and J. Podolsk\'{y}, {\it Exact Space-Times in Einstein's General Relativity} (Cambridge University Press, Cambridge, 2009).
\bibitem{AichelburgSexl:1971} P. C. Aichelburg and R. U. Sexl, Gen. Rel. Grav. {\bf 2}, 303 (1971).
\bibitem{FerrariPendenza:1990}V. Ferrari and P. Pendenza, Gen. Rel. Grav. {\bf 22}, 1105 (1990).
\bibitem{LoustoSanchez:1992}C. O. Loust\'{o} and N. S\'{a}nchez, Nucl. Phys. B  {\bf 383}, 377 (1992).
\bibitem{HottaTanaka:1993}M. Hotta and M. Tanaka, Class. Quant. Gravity {\bf 10}, 307 (1993).
\bibitem{BalasinNachbagauer:1995}H. Balasin and H. Nachbagauer, Class. Quant. Gravity {\bf 12}, 707 (1995).
\bibitem{BalasinNachbagauer:1996}H. Balasin and H. Nachbagauer, Class. Quant. Gravity {\bf 13}, 731 (1996).
\bibitem{PodolskyGriffiths:1997}J. Podolsk\'{y} and J. B. Griffiths, Phys. Rev. {\bf D 56}, 4756 (1997).
\bibitem{PodolskyGriffiths:1998}J. Podolsk\'{y} and J. B. Griffiths, Phys. Rev. {\bf D 58}, 124024 (1998).
\bibitem{Podolsky:2002}J. Podolsk\'{y}, in {\it Gravitation: Following the Prague Inspiration}, ed. by O. Semer\'{a}k, J. Podolsk\'{y}, and M. \v{Z}ofka (World Scientific, Singapore, 2002), pp. 205--246, gr-qc/0201029.
\bibitem{BarrabesHogan:2003}C. Barrab\'{e}s and P. A. Hogan, {\it Singular Null Hypersurfaces in General Relativity} (World Scientific, Singapore, 2003).
\bibitem{Bonnor:1970b} W. B. Bonnor, Int. J. Theor. Phys. {\bf 3}, 257 (1970).
\bibitem{Griffiths:1972} J.~B. Griffiths, Int. J. Theor. Phys. {\bf 5}, 141 (1972).
\bibitem{FrolovFursaev:2005}V. P. Frolov and D. V. Fursaev, Phys. Rev. {\bf D71}, 104034 (2005).
\bibitem{FrolovIsraelZelnikov:2005}V. P. Frolov, W. Israel, and A. Zelnikov, Phys. Rev. {\bf D72}, 084031 (2005).
\bibitem{PravdaPravdovaColeyMilson:2002}V. Pravda, A. Pravdov\'{a}, A. Coley, and R. Milson, Class. Quant. Grav. {\bf 19}, 6213 (2002).
\bibitem{Page:2009}D. N. Page, Class. Quant. Grav. {\bf 26}, 055016 (2009).
\bibitem{ColeyGibbonsHervikPope:2008} A.~A. Coley, G.~W. Gibbons, S. Hervik, and C.~N. Pope., Class. Quant. Grav. {\bf 25},  145017 (2008).
\bibitem{FrolovZelnikov:2006}V. P. Frolov and A. Zelnikov, Class. Quant. Grav. {\bf 23}, 2119 (2006).
\bibitem{FrolovZelnikov:2005}V. P. Frolov and A. Zelnikov, Phys. Rev. {\bf D72}, 104005 (2005).
\bibitem{KadlecovaKrtous:2010} H. Kadlecov\'{a} and P. Krtou\v{s}, Phys. Rev. {\bf D82}, 044041 (2010).
\bibitem{Siklos:1985}S. T. C. Siklos, in {\it Galaxies, Axisymmetric Systems and Relativity}, ed. by M. A. H. MacCallum (Cambridge University Press, Cambridge, 1985), pp. 247--274.
\bibitem{Podolsky:1998} J. Podolsk\'{y}, Class. Quant. Grav. {\bf 15}, 719 (1998).
\bibitem{ColeyHervikPelavas:2006} A. A. Coley, S. Hervik, and N. Pelavas, Class. Quant. Grav. {\bf 23}, 3053 (2006).
\bibitem{ColeyHervikPelavas:2008}A. A. Coley, S. Hervik, and N. Pelavas, Class. Quant. Grav. {\bf 25}, 025008 (2008).
\bibitem{ColeyHervikPelavas:2009}A. A. Coley, S. Hervik, and N. Pelavas, Class. Quant. Grav. {\bf 26}, 025013 (2009).
\bibitem{CaldarelliKlemmZorzan:2007}M.~M. Caldarelli, D. Klemm, and E. Zorzan, Class. Quant. Grav. {\bf 24}, 1341 (2007).
\bibitem{FrolovLin:2006} V.~P. Frolov and F.-L. Lin, Phys. Rev. {\bf D 73}, 104028 (2006).
\bibitem{KadlecovaZelnikovKrtousPodolsky:2009} H. Kadlecov\'{a}, A. Zelnikov, P. Krtou\v{s}, and J. Podolsk\'{y}, Phys. Rev. {\bf D80}, 024004 (2009).
\bibitem{PodolskyZofka:2009} J. Podolsk\'{y} and M. \v{Z}ofka, Class. Quant. Grav. {\bf 26}, 105008 (2009).
\bibitem{ColeyEtal:2009} A.~A. Coley, S. Hervik, G. Papadopoulos and N. Pelavas, Class. Quant. Grav. {\bf 26}, 105016 (2009).
\bibitem{KrtousPodolsky:inprep}P. Krtou\v{s} and J. Podolsk\'{y}, in preparation.
\bibitem{Ficken:1939}F. A. Ficken, Ann. Math. {\bf 40}, 892 (1939).
\bibitem{CardosoDiasLemos:2004} V. Cardoso, \'{O}.~J.~C. Dias, and J.~P.~S. Lemos, Phys. Rev. {\bf D70}, 024002 (2004).
\bibitem{OrtaggioEtal:2008} M. Ortaggio, J. Podolsk\'{y}, and M. \v{Z}ofka, Class. Quant. Grav. {\bf 25}, 025006 (2008).




\end{thebibliography}

\end{document}